\documentclass[graybox, nosecnum, natbib]{svmult}

\usepackage{amsfonts}
\usepackage{mathptmx}       
\usepackage{helvet}         
\usepackage{courier}        
\usepackage{type1cm}        
\usepackage{makeidx}         
\usepackage{graphicx}        
\usepackage{multicol}        
\usepackage[bottom]{footmisc}
\usepackage{hyperref}        
\usepackage{soul}            
\usepackage{braket}
\usepackage{amsmath}
\usepackage{subeqnarray}
\usepackage{fancyhdr}

\newcommand{\gras}[1]{\mathbf{#1}}  
\newcommand{\mydef}[1]{{\em #1}}  

\makeindex             

\begin{document}
\title*{Microscopic Theory of Nuclear Fission}
\author{Nicolas Schunck \thanks{corresponding author}}
\institute{Nicolas Schunck \at Nuclear and Chemical Sciences Division, 
Lawrence Livermore National Laboratory, Livermore, California 94551, USA, 
\email{schunck1@llnl.gov}}

\maketitle

\abstract{Nuclear fission represents the ultimate test for microscopic theories 
of nuclear structure and reactions. Fission is a large-amplitude, 
time-dependent phenomenon taking place in a self-bound, strongly-interacting 
many-body system. It should, at least in principle, emerge from the complex 
interactions of nucleons within the nucleus. The goal of microscopic theories 
is to build a consistent and predictive theory of nuclear fission by using as 
only ingredients protons and neutrons, nuclear forces and quantum many-body 
methods. Thanks to a constant increase in computing power, such a goal has 
never seemed more within reach. This chapter gives an overview both of the set 
of techniques used in microscopic theory to describe the fission process and of 
some recent successes achieved by this class of methods.}

\renewcommand{\headrulewidth}{0pt}
\thispagestyle{fancy}
\fancyhf{}
\rhead{LLNL-JRNL-830606}
\lhead{}

\section{Introduction}

Atomic nuclei can sometimes, either spontaneously or after bombardment by 
external particles, split into two or more fragments. This fission process was 
discovered experimentally in late 1938 by Hahn and Strassman \cite{Hahn1938,
Hahn1939}. Barely six months later, Bohr and Wheeler, building on a critical 
insight by Lise Meitner \cite{Meitner1939}, proposed the first model of the 
process \cite{Bohr1939}. Since then considerable effort has been expended to 
understand this phenomenon. In the past few years, there has been a renaissance 
in fission studies and, as a result, several review articles and textbooks have 
been published on this subject \cite{Krappe2012,Schunck2016,Andreyev2017,
Schmidt2018,Younes2019,Younes2021}. This chapter will focus on the most recent 
developments in building a microscopic theory of fission. Therefore, the 
interested reader is advised to consult the aforementioned references to gain a 
more complete picture of this fascinating problem.

\subsection{Fission Observables}

The complexity of the fission process translates into a large number of 
observables, that is, quantities that can be measured experimentally and, in 
principle, computed theoretically. One may classify these observables into 
three main categories. 

The first category corresponds to the probability that fission takes place in a 
nucleus given a set of initial conditions. The case of spontaneous fission is 
the simplest one: the initial conditions are nothing but the number of protons
and neutrons of the nucleus, which is assumed to be in its ground state, and 
the probability of fission is encoded in the spontaneous fission half life 
$\tau_{\rm SF}$. The case of induced fission is much more involved. First, the 
initial conditions now include the full characteristics of the incident 
particle such as its identity (neutron, photon, proton, $\alpha$ particle, 
etc.), its energy and possibly its quantum numbers, but also the 
characteristics of the target nucleus which may or may not be in its ground 
state. The collision of the incident particle with the target nucleus may take 
many different forms: elastic scattering (the neutron is simply deflected 
without transfer of energy), inelastic scattering (deflection with transfer of 
energy to the target), or capture. Even if the neutron is captured, there is no
guarantee that the resulting excited nucleus will fission. It may decay by 
emitting one or several neutrons (noted as the $(n,xn)$ channel); it may emit a 
series of photons, the $(n,\gamma)$ channel; it may beta decay, the $(n,\beta)$ 
channel; or it may fission, the $(n,f)$ channel. This list is not exhaustive 
and many other channels are potentially available. How the competition between 
these different channels affects the probability of fission is encoded in what 
called the fission cross section. Both the spontaneous fission half life and 
the fission cross section can be measured, at least in some nuclei, with high 
precision.

The second category of observables corresponds, roughly speaking, to the 
outcome of the deformation process that leads the nucleus to its breaking 
point, or scission point. At scission, the initial fission fragments have been 
formed. Each fragment is characterized by its number of protons $Z$ and number 
of neutrons $N$, its excitation energy $E^{*}$, its spin distribution $p(J)$ 
and its level density $\rho(E)$. When a sample of material is irradiated every 
fission event that occurs is different: there is a probability that a fragment 
$(Z,N)$ is formed and this probability is encoded in what is called the primary 
fission fragment distribution. It is customary to use the notation $Y(Z,A)$ to 
refer to the probability (normalized to 200) that a fragment with charge $Z$ 
and total mass $A$ is formed. Marginal distributions $Y(Z)$ (the charge 
distribution) and $Y(A)$ (the mass distribution) can be obtained by integrating 
$Y(Z,A)$ over $A$ and $Z$, respectively. It is essential to realize that all 
the aforementioned quantities cannot be measured because of the extremely short 
time scale at which fission takes place, of the order of $\tau_{f} \approx 
10^{-19}$ s: they must be computed by theoretical models. These models must be 
predictive since the results of the calculations cannot be directly compared 
with data. In fact, the predictions of $Y(Z,A)$, $E^{*}$, $p(J)$, etc., will 
serve as inputs to the statistical reaction theory codes that model the 
deexcitation of the fragments.

This deexcitation phase yields a third class of fission observables, which 
include the characteristics of all the particles that can be emitted. The 
first, prompt emission phase involves mostly neutrons and photons ($\gamma$ 
rays). Observables associated with these particles include the mean number of 
neutrons emitted, $\bar{\nu}$, the average energy of the neutron, the mean 
number of $\gamma$ emitted (called the photon or $\gamma$ multiplicity), 
$N_{\gamma}$, the angular correlations between the neutrons, the photons, or 
the neutrons and the photons, etc. After the prompt emission phase, a number of 
$\beta$ decays may take place. They will result in the emission of electrons 
and antineutrinos.

\subsection{Physics Concepts}

The ultimate goal of any microscopic approach to fission is to achieve a 
description of this process based only on our knowledge of in-medium nuclear 
forces and quantum many-body methods. In an ideal world, this description would 
be based on solving the following equation
\begin{equation}
\ket{\Psi}_{\rm f} = e^{-\frac{it}{\hbar}\hat{H}} \ket{\Psi}_{\rm i} ,
\label{eq:evolution}
\end{equation}
where $\ket{\Psi}_{\rm i}$ and $\ket{\Psi}_{\rm i}$ are, respectively, the 
initial state of the nucleus before fission and the final state after the 
process has completed, and $\hat{H}$ is the nuclear many-body Hamiltonian. This 
naive approach is faced with formidable challenges:
\begin{itemize}
\item The nuclear many-body Hamiltonian $\hat{H}$ is not known exactly. Our 
best current model of it is based on chiral effective field theory 
\cite{Weinberg1990,Machleidt2016,Hammer2020}. The characteristic features of 
$\hat{H}$, most notably the short range of nuclear fores and the presence of 
3-, 4- and generally $N$-body interaction terms, make determining the nuclear 
wave functions an extremely challenging problem even in the lightest nuclei 
\cite{Barrett2013}. In recent years, the scope of {\it ab initio} methods -- 
very loosely defined here as the set of methods trying to solve the nuclear 
many-body problem with a microscopic Hamiltonian $\hat{H}$ such as the one from 
chiral EFT -- has extended tremendously thanks to the introduction of powerful 
many-body techniques and the constant improvement of supercomputers 
\cite{Hagen2014,Carlson2015,Hergert2016,Stroberg2019}. Unfortunately, these 
methods cannot reach the region of actinides and superheavy nuclei where 
fission is relevant and remain largely focused on static properties of nuclei. 
Furthermore, the precision of such calculations is not sufficient yet.
\item The very definition of the initial and final states in
(\ref{eq:evolution}) is far from trivial. In all types of fission, the final 
state involves not a single nucleus, but a pair of fragments together with a 
number of emitted particles (neutrons, photons, possibly electrons and 
anti-neutrinos if $\beta$-decay is included). Even the determination of the 
initial wave function $\ket{\Psi}_{\rm i}$ is a challenge. In spontaneous 
fission, it is associated with the ground state of the nucleus and might thus, 
in a near future, be within reach of {\it ab initio} methods. In the case of 
induced, fission, however, it entails characterizing the absorption of an 
incoming particle (neutron or photon, typically) by a heavy target nucleus for 
a broad range of incident energies. Extending the methods used in light-ion 
reactions \cite{Baroni2013a} to such heavy nuclei is not an easy task.
\end{itemize}
For these reasons, the adjective ``microscopic'' in this chapter will take a 
rather narrow definition. It will refer to the set of methods globally known 
as \mydef{energy density functional} theory (EDF) \cite{Schunck2019}. Most of 
the techniques developed in the EDF approach are about finding good 
approximations to the nuclear many-body wave function without having to deal 
with determining the spectrum of the true nuclear many-body Hamiltonian 
$\hat{H}$. By construction, the EDF theory thus contains a higher degree of 
phenomenology than {\it ab initio} approaches. At the same time, it is still a 
many-body theory of the nucleus, by contrast with, e.g., the liquid drop model.

Ever since the very first model of the process by Bohr and Wheeler 
\cite{Bohr1939}, fission has always been thought of as an extreme deformation 
process. This concept is very easily built in phenomenological approaches based 
on the liquid drop model. However, since the nuclear Hamiltonian $\hat{H}$ is 
invariant under rotations, the nuclear wave functions $\ket{\Psi}$ should also 
be rotationally invariant and characterized by good angular momentum. Clearly, 
this is not compatible with the concept of deformation. This could pose 
interesting conceptual difficulties if fission were described with 
{\it ab initio} methods; in the context of the EDF approach, deformation is in 
fact a natural consequence of a phenomenon called \mydef{spontaneous symmetry 
breaking} \cite{Duguet2014,Schunck2019}. When solving the Hartree-Fock or 
Hartree-Fock-Bogoliubov equation of the single-reference EDF theory (see 
below), solutions that break some or all of the symmetries of the nuclear 
Hamiltonian can lower the energy of the system. This effect, also known as the 
Jahn-Teller effect, is present in many fields of physics, from nuclei to 
electronic systems to quantum field theory; see \cite{Nazarewicz1994} for a 
discussion in the context of nuclear physics. 

The concept of spontaneous symmetry breaking is also closely related to the one 
of \mydef{collective variables}. In molecular systems, for example, the 
separation of all degrees of freedom into slow, collective ones associated with 
the coordinates $R_i$ of the ions and the fast, intrinsic degrees of freedom 
related to the electrons is straightforward. In nuclear systems, this 
distinction is less clear but still applies. As recognized a long time ago, 
the low-lying spectrum of nuclei contains sequences of excited states that 
suggest strong collective behavior of the nucleus as whole, akin to collective 
vibrations and rotations \cite{Bohr1998,Ring2004}. The order parameters of each 
of the symmetry groups of the Hamiltonian can in fact provide an initial set of 
collective variables. For example, the multipole moments of the one-body 
density 
\begin{equation}
\rho_{\lambda\mu} 
= \braket{\hat{Q}_{\lambda\mu}\rho}
= \int d^{3}\gras{r}\, Q_{\lambda\mu}(\gras{r})\rho(\gras{r})
\end{equation}
are the (norm of) the order parameter associated with the group of spatial 
rotations. Note that the deformed Hartree-Fock-Bogoliubov solutions can still 
be characterized by many, lower-order point symmetries \cite{Dobaczewski2000a,
Dobaczewski2000b}: these provide additional collective variables controlling 
only very specific features of the nuclear shape. One of the key assumptions of 
most fission models, whether or not they are microscopic, is that fission is 
driven to a large extent by a few important collective variables. This 
assumption is supported {\it ex post} by the excellent agreement with 
experimental data that models achieve for nearly all the fission observables 
that we defined above. In this chapter, collective variables will be denoted by 
the generic vector $\gras{q} = (q_1,\dots,q_N)$. With each collective variable 
$q_i$, we can associate a collective momentum 
$p_i = i\hbar\partial/\partial q_i$. The set of vectors $\mathcal{P} = 
(\gras{q},\gras{p})$ defines the classical phase space for the collective 
motion of the nucleus. The number of $N$ of collective variables and their 
definition are very model-dependent and guided mostly by a comparison of 
theoretical predictions with experimental data.

On the other hand, the interplay between collective and non-collective, 
sometimes called intrinsic, degrees of freedom leads us to the concept of 
\mydef{dissipation} \cite{Norenberg1983}. During its motion through its 
classical phase space $\mathcal{P}$, some of the collective energy of the 
nucleus may be converted into non-collective excitations. This mechanism is in 
fact built in the so-called dissipation tensor of the Langevin equation used to 
simulate fission events as classical trajectories in the multi-dimensional 
potential energy surface of the nucleus \cite{Abe1996,Frobrich1998,Sierk2017}. 
Dissipation results in a loss of collective momentum and may have a large 
impact on some fission observables, especially the properties of the fragments 
when they are formed. As we will see below, time-dependent density functional 
theory simulations suggest that the degree of dissipation is very large, which 
results in a very slow collective motion \cite{Tanimura2015,Bulgac2016,
Bulgac2019a}. Note that in such calculations dissipation is mostly caused by  
\mydef{diabatic} single-particle or quasi-particle (when pairing is included) 
motion: changes in the value of collective variables cause (quasi)particles 
levels of same symmetry to cross in such a way that the system at deformation 
$q' > q$ can be viewed, in practice, as a complex particle-hole excitation of 
the system at $q$. In addition to this mechanism, dissipation can also take 
place amongst collective modes themselves \cite{Younes2012}.

One of the most spectacular manifestations of the quantum-mechanical nature of 
fission is the phenomenon of spontaneous fission: without any stimulus from an 
external source, atomic nuclei in theirs ground state spontaneously break apart. 
Spontaneous fission is made possible by the \mydef{tunneling} effects of 
quantum mechanics, namely the ability of systems to escape a stable potential 
energy well (with some probability) \cite{Messiah1961}. In classes of quantum 
mechanics, examples of tunneling often involve a single particle of mass $m$ in 
a one-dimensional potential well (and for educational purposes, such potentials 
have often analytical forms such as a square well); in atomic nuclei, the 
situation is a bit more complex for two reasons. First, the potential well is 
defined in the collective space spanned by the collective coordinates 
$\gras{q}$ introduced above. It is thus multi-dimensional and certainly not 
analytical. Most importantly, the ``mass'' of the system in that collective 
space is not the atomic mass of the nucleus or the nuclear binding energy. It 
is in fact better thought of as the response of the nucleus to a change in 
collective variable and is not a scalar but a rank-2 tensor. The role of this 
collective mass tensor in determining tunneling probabilities is very important 
as we will see in the next section.

The last important, and somewhat elusive, concept in fission theory is the one 
of \mydef{scission}. Qualitatively, scission is the moment where the nucleus 
breaks into two (or more) fragments. The characteristics of the fission 
fragments right at the moment they are formed -- number of protons and 
neutrons, excitation energy, spin, parity -- will determine their subsequent 
decay. However, quantum-mechanical systems do not behave like rigid bodies or 
classical fluids and do not separate as neatly. Fragments may remain entangled, 
at least for some time, even after being separated \cite{Younes2011}. In fact, 
the system that EDF methods describe is always the fissioning nucleus: by 
construction, the many-body wave function of that system must remain 
anti-symmetrized at all time, which induces a form of (possibly) spurious 
entanglement between the fragments. 

\begin{figure}[!ht]
\centering
\includegraphics[width=0.8\textwidth]{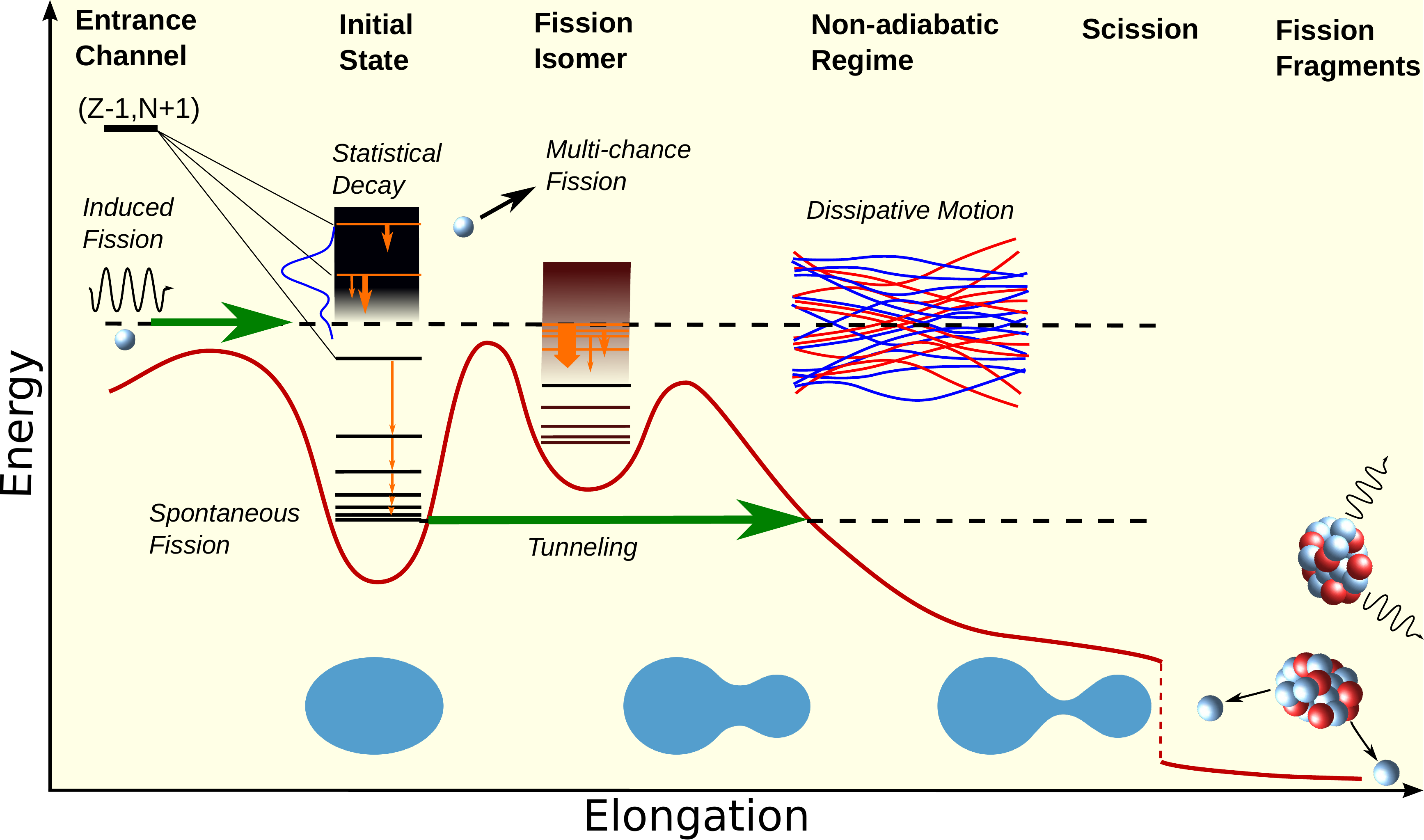}
\caption{Schematic illustration of the main physics concepts at play in 
fission. Potential energy surfaces (in one dimension like here: a potential 
energy curve) play a key role in determining the spontaneous fission 
probabilities, as well as the identities of the fission fragments and their 
distribution. The level structure above the ground state (first minimum) and 
the fission isomer (second minimum) will impact the fission cross sections. 
What is called the descent from saddle to scission involves a very slow, 
diabatic motion characterized by multiple level crossings and conversion 
between collective and intrinsic energy.
}
\label{fig:physics}
\end{figure}

Figure \ref{fig:physics} summarizes the main physics concepts relevant in 
fission theory. The figure represents the variation of the total energy as a 
function of the elongation, with several pictorial illustrations of the nuclear 
shape along the way. As mentioned above, the very introduction of the notion of 
elongation, that is, deformation, requires invoking the phenomenon of 
spontaneous symmetry breaking at play in EDF theories. It also requires 
specifying a set of collective variables that will quantify this deformation. 
For the sake of simplicity, the figure was made by considering a single 
collective variable. The interplay between collective motion and intrinsic 
excitation, aka the dissipative motion of the system, is especially relevant 
beyond the second barrier. The point of scission, where the two fragments are 
formed, is represented here as a sharp drop in total energy.

\section{Theoretical Models}

As mentioned above, most microscopic approaches to fission are based on the 
nuclear energy density functional (EDF) framework \cite{Schunck2019}. The term 
energy density functional methods encompasses both a set of general many-body 
techniques such as the Hartree-Fock, BCS and Hartree-Fock-Bogoliubov theories, 
the QRPA or the generator coordinate method, and an approximation for the 
in-medium nuclear forces inside the nucleus. The many-body techniques offer an 
ansatz for the many-body wave function representing the nucleus, while the 
effective Hamiltonian, or energy functional, modeling nuclear forces, ensures 
that said ansatz ``works'' and allows reproducing nuclear observables such as 
binding energies, excited states, etc. 

\subsection{Energy Density Functional Theory}

In all recent applications, the basic building block of the EDF approach is the 
Hartree-Fock-Bogoliubov (HFB) theory \cite{Valatin1961}, which provides a 
general method to determine an approximation of both the ground state and 
excited states for a many-body quantum system described by a product-state wave 
function. The HFB theory has been presented in great details in the review 
articles \cite{Valatin1961,Mang1975,Bender2003} as well as in several textbooks 
\cite{Blaizot1985,Ring2004,Schunck2019} and below we will only give a very brief 
summary.

\subsubsection{Hartree-Fock-Bogoliubov Theory}

{\bf General Formalism -}
The HFB theory is most conveniently explained in the language of second 
quantization. We begin with a basis of single-particle states represented by a 
set of fermion creation and annihilation operators $(c_{m}^{\dagger},c_{m})$ 
with $m=1,\dots,+\infty$. Note that in practical applications, the 
single-particle basis is truncated to have only a finite number of terms $p$. 
This basis can be, for instance, associated with the eigenstates of the 
harmonic oscillator. The Bogoliubov transformation introduces a new set 
$(\beta_{\mu}^{\dagger},\beta_{\mu})$ of fermion operators, 
\begin{eqnarray}
\beta_{\mu} & = \displaystyle\sum_{m} \left[ U^{\dagger}_{\mu m}\,c_{m} + V^{\dagger}_{\mu m}\,c_{m}^{\dagger} \right] ,
\medskip\\
\beta_{\mu}^{\dagger} & = \displaystyle\sum_{m} \left[ V^{T}_{\mu m}\,c_{m} + U^{T}_{\mu m}\,c_{m}^{\dagger} \right].
\end{eqnarray}
where $U$ and $V$ form the Bogoliubov matrix
\begin{equation}
\mathcal{W} = \left( \begin{array}{cc}
U & V^{*}  \\
V & U^{*}
\end{array}\right) .
\end{equation}
The operators $\beta_{\mu}^{\dagger}$ and $\beta_{\mu}$ are the quasiparticle 
(q.p.) operators. In the HFB theory, one assumes that the many-body wave 
function can be represented as a product state of quasiparticle operators,
\begin{equation}
\ket{\Phi} = \prod_{\mu} \beta_{\mu}\ket{0},
\label{eq:HFB_ansatz}
\end{equation}
where $\ket{0}$ is the particle vacuum. Quasiparticles can be interpreted as 
excitations of the system: $\beta_{\mu}^{\dagger}\ket{0}$ creates an excited 
state on top of the vacuum. The ansatz (\ref{eq:HFB_ansatz}) thus translates 
the fact that the HFB ground state is a zero-excitation state.

Given the form (\ref{eq:HFB_ansatz}) of the HFB wave function, the next step is 
to compute the total energy. For the sake of simplicity, we will only discuss 
the case where the energy is calculated as the expectation value of an 
effective Hamiltonian $\hat{H}_{\rm eff}$, 
\begin{equation}
E = \frac{\braket{\Phi|\hat{H}_{\rm eff} | \Phi}}{\braket{\Phi|\Phi}} .
\end{equation}
We refer the interested readers to more complete presentations for the case 
where the energy is derived from an energy functional \cite{Schunck2019}. When 
the effective Hamiltonian reads $\hat{H} = \hat{T} + \hat{V}$ with $\hat{V}$ 
containing only two-body effective forces, the total energy becomes
\begin{equation}
E[\rho,\kappa,\kappa^{*}] = 
\sum_{ij} t_{ij} \rho_{ji}
+
\frac{1}{2} \sum_{ijkl} \bar{v}_{ijkl}\rho_{lj}\rho_{ki}
+
\frac{1}{4}\sum_{ijkl} \bar{v}_{ijkl} \kappa^{*}_{ij}\kappa_{kl}.
\label{eq:HFB_energy}
\end{equation}
where $\rho$ is the one-body density matrix and $\kappa$ the pairing tensor 
defined as
\begin{equation}
\rho_{mn} =\frac{\braket{\Phi|c_{n}^{\dagger}c_{m}|\Phi}}{\braket{\Phi|\Phi}},\qquad
\kappa_{mn} = \frac{\braket{\Phi|c_{n}c_{m}|\Phi}}{\braket{\Phi|\Phi}},\qquad
\kappa_{mn}^{*} = \frac{\braket{\Phi|c_{m}^{\dagger}c_{n}^{\dagger}|\Phi}}{\braket{\Phi|\Phi}} ,
\end{equation}
and the matrix elements of the operators are $t_{ij} = \braket{i|\hat{T}|j}$ 
(kinetic energy) and $v{ijkl} = \braket{ij|\hat{V}|kl}$ (two-body potential). 
The notation $\bar{v}_{ijkl} = v_{ijkl} - v_{ijlk}$ indicate the 
antisymmetrization  of all matrix elements. By construction, the degrees of 
freedom in the HFB theory are the matrix elements of the Bogoliubov 
transformation or, equivalently, those of the one-body density matrix and 
pairing tensor. They are determined by requiring that the total energy 
(\ref{eq:HFB_energy}) be minimum with respect to their variations. This leads 
to the HFB equation, which takes the form of the commutator
\begin{equation}
\big[ \mathcal{H}, \mathcal{R} \big] = 0 ,
\label{eq:HFB_eq}
\end{equation}
where the HFB matrix $\mathcal{H}$ and generalized density $\mathcal{R}$ read
\begin{equation}
\mathcal{H} = \left( \begin{array}{cc}
h   & \Delta  \\
-\Delta^{*} & -h^{*}
\end{array}\right),
\qquad
\mathcal{R} =
\left( \begin{array}{cc}
\rho   & \kappa \\
-\kappa^{*} & 1 - \rho^{*}
\end{array}\right) ,
\label{eq:HFB_matrices}
\end{equation}
with $h_{ij} = t_{ij} + \sum_{kl} \bar{v}_{ikjl}\rho_{lk}$ is the mean field 
and $\Delta_{ij} = \frac{1}{2}\sum_{kl} \bar{v}_{ijkl}\kappa_{kl}$ the pairing 
field. Solving (\ref{eq:HFB_eq}) determines completely the generalized density, 
hence $\rho$ and $\kappa$. Physical observables can be computed by taking the 
trace of the corresponding operator with the densities. In the simplest case of 
one-body operators $\hat{F}$, this means: $\braket{\hat{F}} = 
\mathrm{Tr} (\hat{F}\rho) = \sum_{ij} F_{ij}\rho_{ji}$. The expectation value 
of two- and $N$-body operators can be computed thanks to the Wick theorem 
\cite{Ring2004,Schunck2019}.

{\bf Constraints - }
As mentioned earlier in the introduction, the theoretical description of 
fission involves mapping out the variations of the total energy as a function 
of the deformation of the fissioning nucleus. In phenomenological models based 
on the macroscopic-microscopic method, this is achieved by parameterizing the 
nuclear shape; in the HFB theory, this requires seeking solutions of the HFB 
equation (\ref{eq:HFB_eq}) with constraints on the expectation value of 
operators associated with the nuclear shape. The simplest example of such 
operators are the multipole moments (computed in the intrinsic frame of 
reference): $\hat{Q}_{\lambda\mu} = c_{\lambda\mu}r^{\lambda}Y_{\lambda\mu}$, 
where $c_{\lambda\mu}$ is a normalization constant and $Y_{\lambda\mu}$ are the 
spherical harmonics \cite{Varshalovich1988}. Let us note 
$\{ \hat{F}_{a} \}_{a=1,\dots,N}$ the set of $N$ operators on which expectation 
value we seek to impose a constraint. This can be achieved by the standard 
method of Lagrange parameters: instead of minimizing the total energy 
$E[\rho,\kappa,\kappa^*]$, we minimize instead 
$\mathcal{E} = E - \sum_{a} (\braket{\hat{F}_a} - \bar{F}_a)$, where 
$\bar{F}_a$ is the target value for the constraint and $\braket{\hat{F}_a} =
\mathrm{Tr}(\hat{F}_a\rho)$ the expectation value (here assuming a one-body 
operator only). It is straightforward to see that the HFB matrix is modified 
and becomes 
\begin{equation}
\mathcal{H} = \left( \begin{array}{cc}
h - \sum_a F_{a}  & \Delta  \\
-\Delta^{*} & -h^{*} + \sum_a F^{*}_{a}
\end{array}\right),
\end{equation}
where $F_a$ is now the matrix of the operator $\hat{F}_a$ in the 
single-particle basis. Even without specific constraints on the nuclear shape, 
this technique is needed to ensure the conservation of particle number in the 
HFB equation: choosing $\hat{F}_a \equiv \hat{N}$ ensures that HFB solution 
(\ref{eq:HFB_ansatz}) has the right number of particles on average. 

There are two main techniques to solve the HFB equation (\ref{eq:HFB_eq}). As 
well known from basic quantum mechanics, if two operators commute, it is 
possible to find a common eigenbasis. One popular approach to solving the HFB 
equation thus involves recasting (\ref{eq:HFB_eq}) into a non-linear eigenvalue 
problem which is solved iteratively. Alternatively, one may view the HFB 
equation as a minimization problem. By suitably parameterizing the HFB wave 
function thanks to the Thouless theorem, one can adapt gradient techniques to 
minimize the energy. These two techniques are discussed in details in 
\cite{Schunck2019}.

{\bf Spontaneous symmetry breaking -} 
As mentioned above, the ansatz (\ref{eq:HFB_ansatz}) for the HFB wave function 
implies that this wave function is not an eigenstate of the particle number 
operator $\hat{N} = \sum_{i} c_i^{\dagger}c_i$. Similarly, solving the HFB 
equation (\ref{eq:HFB_eq}) with constraints on the expectation value of multipole 
moments in the intrinsic frame implies that the density matrix $\rho$ is not 
going to be rotationally invariant. Two fundamental symmetries of the nuclear 
Hamiltonian, particle number and rotational invariance, are thus {\it broken} 
in the HFB theory. This spontaneous symmetry breaking is a trademark of the 
EDF theory and one of the main reasons why, in practice, this approach provides 
an accurate model for atomic nuclei. As discussed extensively in \cite{Ring2004, 
Duguet2014,Schunck2019}, the symmetry group $\mathcal{G}$ of the nuclear 
Hamiltonian reads
\begin{equation}
\mathcal{G} = T(3) \times SO(3) \times \Pi \times U(1)_{N} \times U(1)_{Z} \times SU(2)
\end{equation}
where $T(3)$ is the Abelian Lie group of space translations associated with the 
conservation of the linear momentum; $SO(3)$ is the non-Abelian Lie group of 
space rotations associated with the conservation of angular momentum; $\Pi$ is 
the Abelian, finite and discrete group associated with reflection symmetry; 
$U(1)_{X}$ is the Abelian Lie group associated with the conservation of the 
number of particles ($X=Z$ for protons, $X=N$ for neutrons) and $SU(2)$ is the 
group associated with isospin symmetry. Saying that $\mathcal{G}$ is the 
symmetry group of the nuclear Hamiltonian means that for any element 
$g\in\mathcal{G}$ and the relevant unitary operator $\hat{U}(g)$, we have: 
$[\hat{U}(g),\hat{H}] = 0$. The key feature of the EDF approach is to seek 
solutions to the HFB equation that can break any of these symmetries. By doing 
so, we explore a variational space that is richer and, as a consequence, can 
lower the total energy: the product state (\ref{eq:HFB_eq}) contains 
correlations that would not be present otherwise and combined with suitably 
chosen effective Hamiltonian $\hat{H}_{\rm eff}$ gives a realistic 
approximation of the nuclear wave function. It is important to realize that the 
microscopic description of fission is entirely built upon this concept and the 
related mathematical framework. As of now, we do not have an alternative 
description of the phenomenon rooted, e.g., in the nuclear shell model or ab 
initio methods, where symmetries are never broken.

{\bf Energy functionals -}
The previous paragraphs summarized important general concepts of the EDF 
theory. In practical applications, one must specify the effective Hamiltonian 
$\hat{H}_{\rm eff}$ or, alternatively, the energy functional 
$E[\rho,\kappa,\kappa^*]$ in order to perform calculations. The Skyrme and 
Gogny energy functionals are among the most commonly used models for nuclear 
forces \cite{Bender2003,Stone2007,Robledo2019,Schunck2019}. These functionals 
are derived from non-relativistic, two-body effective Hamiltonians. The Skyrme 
Hamiltonian is a local zero-range potential, i.e., of the type 
$\hat{V}(\gras{r}_1,\gras{r}_2)\delta(\gras{r}_1-\gras{r}_2)$ with $\gras{r}_1$ 
and $\gras{r}_2$ the coordinates of the two nucleons; the Gogny Hamiltonian is a 
local, finite-range potential of the type $\hat{V}(\gras{r}_1,\gras{r}_2)$ 
where the dependency on $\gras{r}_1$ and $\gras{r}_2$ is proportional to 
$\exp\big[(\gras{r}_1 - \gras{r}_2)^2/\mu^2 \big]$ with $\mu$ a typical length 
scale. Each of these Hamiltonian contains a central part, a density-dependent 
part that mocks up the effect of three-body forces and a spin-orbit part. In 
addition, both effective Hamiltonians contain a Coulomb potential. While the 
form (\ref{eq:HFB_energy}) suggests that the same potential should be used to 
compute both the mean field energy (proportional to $\rho\rho$) and the pairing 
energy (proportional to $\kappa\kappa^*$), practitioners tend to adopt the 
following, more flexible (and less consistent) formula
\begin{equation}
E[\rho,\kappa,\kappa^{*}] = 
\sum_{ij} t_{ij} \rho_{ji}
+
\frac{1}{2} \sum_{ijkl} \bar{v}_{ijkl}^{\rho\rho}\rho_{lj}\rho_{ki}
+
\frac{1}{4}\sum_{ijkl} \bar{v}_{ijkl}^{\kappa\kappa} \kappa^{*}_{ij}\kappa_{kl}.
\label{eq:HFB_energy_1}
\end{equation}
where the potential in the mean-field and pairing channel is {\it different}. 
In such cases, the matrix elements $\bar{v}_{ijkl}^{\kappa\kappa}$ of the 
pairing channel are typically computed from a simple pairing functional, e.g., 
a density-dependent zero-range pairing force \cite{Dobaczewski2002}.

\subsubsection{Multi-Reference Energy Density Functional}

While the mechanism of spontaneous symmetry breaking discussed above seems at 
first sight ideal to describe the extreme deformation process that fission is, 
it also leads to the unintended consequences discussed in the introduction: all 
the quantum numbers associated with the broken symmetries of the group 
$\mathcal{G}$ are lost. This means that a deformed nucleus described by an HFB 
wave function of the type (\ref{eq:HFB_ansatz}) cannot be labelled by angular 
momentum $JM$ and parity $\pi$, in addition to not being even associated with a 
fixed number of particles $Z$ and $N$. In fact, the single-reference EDF wave 
function can be written as the {\it wave packet}
\begin{equation}
\ket{\Phi} = \sum_{JM}\sum_{\pi} \sum_{ZN} c(Z,N,J,M,\pi) \ket{\Psi^{ZN}_{JM\pi}}, 
\label{eq:expansion}
\end{equation}
where $\ket{\Psi^{ZN}_{JM\pi}}$ is an eigenstate of the particle number 
operators (both protons and neutrons), of $\hat{J}^{2}$ and $\hat{J}_z$ and of 
the parity operator $\hat{P}$, with $c(Z,N,J,M,\pi)$ the coefficients of the 
expansion that depend on all the relevant quantum numbers. It is worth noting 
that the expansion (\ref{eq:expansion}) also applies to more phenomenological 
theories of fission such as the macroscopic-microscopic model. In fact, it is 
inherent to any theory relying on spontaneous symmetry breaking. In practice, 
the expansion(\ref{eq:expansion}) implies that the potential energy surface 
computed, e.g., at the HFB approximation, does not correspond to the nucleus 
$(Z,N)$ (even if one used constraints to enforce the {\it average} value of 
both $Z$ and $N$)), but to a superposition of different nuclei with different 
numbers of protons and neutrons. Similarly, one could introduce constraints on 
the average value $\braket{\hat{J}}$ of angular momentum to produce potential 
energy surface at different spins, but the resulting surfaces would be, 
strictly speaking, superpositions of surfaces at different $J$.

{\bf Projection Techniques -} Quantum numbers can be recovered by projecting 
HFB solutions on the appropriate subspace of the Hilbert space. Starting with a 
symmetry-breaking HFB wave function $\ket{\Phi(g)}$, where $g$ is the parameter 
corresponding to some some continuous symmetry $S$ of the symmetry group 
$\mathcal{G}$, one can define a formal projector $\hat{P}_S$ as
\begin{equation}
\ket{\Psi} = \hat{P}_S\ket{\Phi(g)} = \int dg' f(g') \hat{R}(g')\ket{\Phi(g)},
\label{eq:proj}
\end{equation}
where $f(g')$ is a weight function and $\hat{R}(g')$ is the unitary 
transformation that changes the ``orientation'' $g$ of the wave function. Both
$f(g')$ and $\hat{R}(g')$ depend on the symmetry group under consideration. For 
example, for the $U(1)_{N}$ group associated with particle number $N$, we have
\begin{equation}
U(1)_{N}: \quad 
\begin{array}[t]{rcl}
\displaystyle \int dg'   & \equiv & \displaystyle \int_{0}^{2\pi}d\varphi, \medskip\\
\displaystyle f(g')      & \equiv & e^{+i\varphi N}, \medskip\\
\displaystyle \hat{R}(g') & \equiv & e^{-i\varphi \hat{N}} .\\
\end{array}
\label{eq:PNP}
\end{equation}
Note that in the expression of $f(g')$, $N$ appears as a scalar while in that 
of $\hat{R}(g')$, it is the operator. Another important example is angular 
momentum where
\begin{equation}
SO(3): \quad 
\begin{array}[t]{rcl}
\displaystyle \int dg'    & \equiv & \displaystyle \frac{2J+1}{16\pi^2}\int_{0}^{2\pi} d\alpha \int_{0}^{\pi} d\beta\sin\beta \int_{0}^{4\pi}d\gamma, \medskip\\
\displaystyle f(g')       & \equiv & D_{MK}^{J*}(\alpha,\beta,\gamma), \medskip\\
\displaystyle \hat{R}(g') & \equiv & e^{-i\alpha\hat{J}_z}e^{-i\beta\hat{J}_y}e^{-i\gamma\hat{J}_z}.
\end{array}
\label{eq:AMP}
\end{equation}
Qualitatively, the effect of the projector is best understood in the case of 
rotations: it takes a given, deformed HFB solution $\ket{\Phi}$, rotates it in 
space by Euler angles $\Omega = (\alpha,\beta,\gamma)$ and combines the results 
to produce a rotationally-invariant wave function $\ket{\Psi}$. Details on the 
formalism, implementation and formal issues associated with projection 
techniques can be found in many review articles and textbooks; see 
\cite{Mang1975,Ring2004,Schunck2019,Bally2021,Sheikh2021} for a short 
selection. In the context of fission theory, projection techniques have been 
applied only in a handful of cases, most notably to study the changes in the 
potential energy surfaces caused by the projection-induced correlation energy 
\cite{Bender2004a,Hao2012,Marevic2020}. One of the most promising applications 
of projection has to do with the determination of the number of particles and 
spin distributions of fission fragments, which we will discuss later.

{\bf Configuration Mixing -} As briefly mentioned earlier, spontaneous symmetry 
breaking allows exploring a richer variational space for the HFB solution 
$\ket{\Phi}$ which, however, remains an independent quasiparticle product state 
of the form (\ref{eq:HFB_ansatz}). Following original ideas by Griffin and 
Wheeler, additional correlations can be encoded in the wave functions by 
performing a configuration mixing of different HFB states \cite{Griffin1957}, 
schematically,
\begin{equation}
\ket{\Psi} = \int d\gras{q}\, f(\gras{q})\ket{\Phi(\gras{q})} ,
\label{eq:GCM_ansatz}
\end{equation}
where $f(\gras{q})$ is a weight function and $\ket{\Phi(\gras{q})}$ is the HFB 
state at point $\gras{q}$ in the collective space. The ansatz 
(\ref{eq:GCM_ansatz}) is very general. In practice, the states 
$\ket{\Phi(\gras{q})}$ are known, e.g., by performing a series of HFB 
calculations, and only the weight functions $f(\gras{q})$ must be determined. 
This is done by inserting (\ref{eq:GCM_ansatz}) into the many-body 
Schr\"odinger equation. This results in what is known as the 
Hill-Wheeler-Griffin equation
\begin{equation}
\int d\gras{q} \big[ \mathcal{H}(\gras{q}',\gras{q}) - \mathcal{E}_{\mu}\mathcal{N}(\gras{q}',\gras{q}) \big]f_{\mu}(\gras{q}) = 0,
\end{equation}
where $\mathcal{E}_{\mu}$ are the energies of the collective states 
$\ket{\Psi_{\mu}}$, $\mathcal{H}(\gras{q}',\gras{q})$ is the energy kernel and 
$\mathcal{N}(\gras{q}',\gras{q})$ is the norm kernel, given by
\begin{align}
\mathcal{N}(\gras{q}',\gras{q}) & = \braket{\Phi(\gras{q})|\Phi(\gras{q}')}, \\
\mathcal{H}(\gras{q}',\gras{q}) & = \braket{\Phi(\gras{q})|\hat{H}|\Phi(\gras{q}')}
\end{align}
This configuration mixing is known in the literature as the generator 
coordinate method (GCM) and has a long history; we refer the interested reader 
to \cite{WaWong1975,Reinhard1987,Bender2003,Ring2004} for further details. We 
will discuss the applications of the GCM framework to fission in the context of 
collective models.

\subsection{Time-Dependent Density Functional Theory}
\label{subsec:tddft}

\begin{figure}[!ht]
\centering
\includegraphics[width=0.8\textwidth]{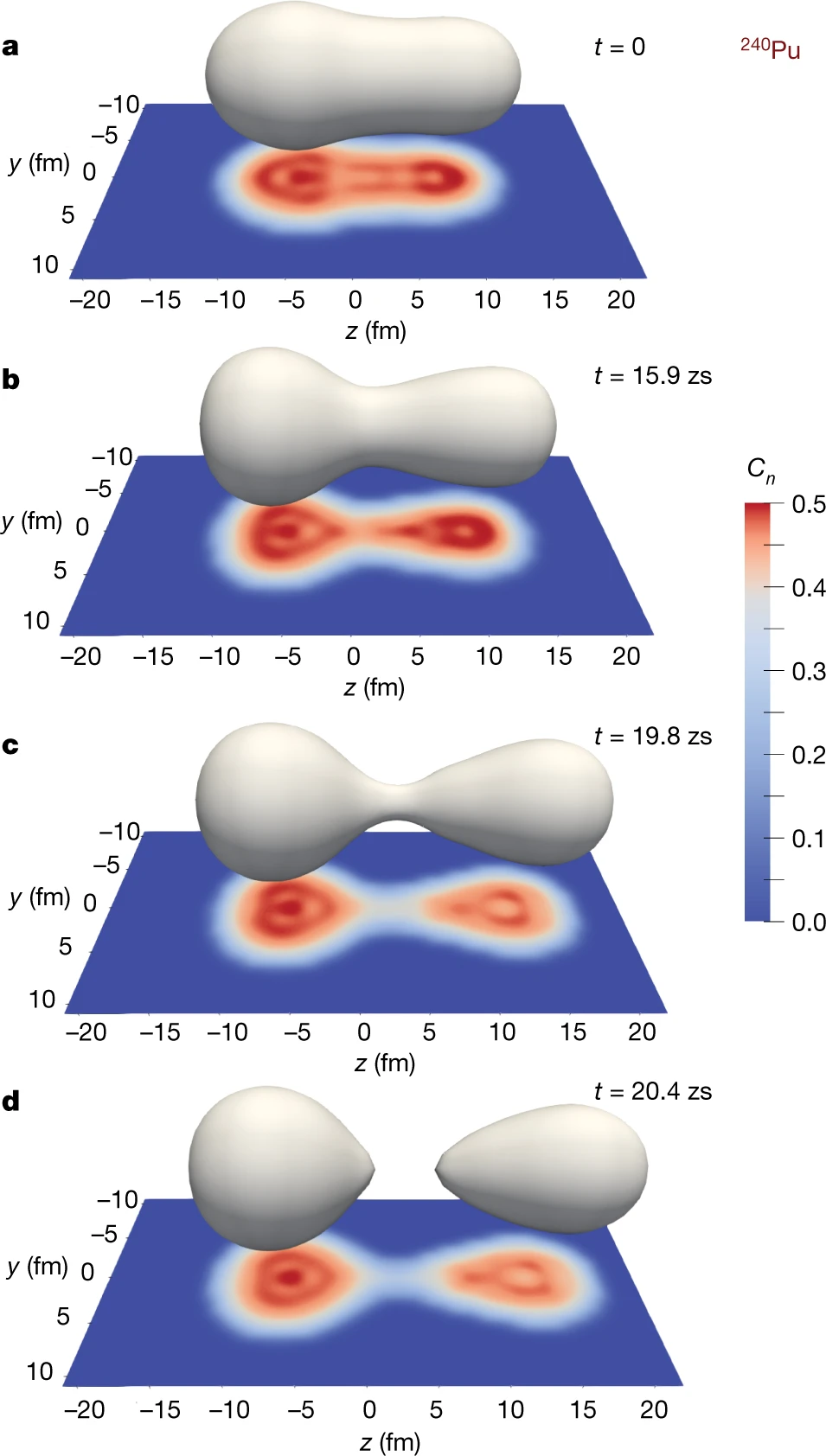}
\caption{Real-time evolution of the one-body density of the $^{240}$Pu 
asymmetric fission from TDHF+BCS calculations. The 3D surface highlights the 
half-saturation density (0.08 fm$^{-3}$) isosurface whereas the projected color
map corresponds to a localization function of the nucleons. 
Figures reproduced with permission from \cite{Scamps2018} courtesy of Scamps;
}
\label{fig:tddft}
\end{figure}

Most of the preceding presentation of the EDF theory was focused on static, 
i.e., time-independent, nuclear properties. The EDF theory can also be extended 
to time-dependent phenomena through what we will refer, somewhat loosely, as 
time-dependent density functional theory (TDDFT) \cite{Bulgac2013a,
Nakatsukasa2016a}. The starting point of TDDFT is the time-dependent many-body 
Schr\"odinger equation,
\begin{equation}
i\hbar \frac{\partial\ket{\Psi(t)}}{\partial t} = \hat{H}\ket{\Psi(t)},
\label{eq:mb_schrod}
\end{equation}
where $\ket{\Psi(t)}$ is the exact, time-dependent many-body wave function of 
the system and $\hat{H}$ the nuclear many-boy Hamiltonian. The next step is to 
adopt an ansatz for the form of $\ket{\Psi(t)}$. In the time-dependent 
Hartree-Fock (TDHF) theory, we would enforce that $\ket{\Psi(t)}$ remains a 
product state of {\it single-particle} states at all times (=a Slater 
determinant); in the time-dependent Hartree-Fock-Bogoliubov (TDHFB) theory, we 
would similarly choose that $\ket{\Psi(t)}$ must be a Bogoliubov vacuum of the 
form (\ref{eq:HFB_ansatz}) at all times (in other words: the operators 
$\beta_{\mu}$ are now time-dependent). The third step is to insert the chosen 
ansatz for the wave function into (\ref{eq:mb_schrod}) and find a method to 
determine either the single-particle wave functions $\varphi_k(t)$ (TDHF) or 
the Bogoliubov matrices $U(t)$ and $V(t)$ (TDHFB). This step is not as 
straightforward as in the static case as shown in \cite{Blaizot1985}; see also 
\cite{Balian1985,Balian1988,Lacroix2010,Lacroix2011} for additional 
discussions. Here, we only recall the basic formula that are relevant for 
fission theory. In TDHF, the application of the variational principle leads to 
the following equation,
\begin{equation}
i\hbar \frac{\partial\rho}{\partial t} = \big[ h[\rho(t)], \rho(t) \big],
\label{eq:tdhf}
\end{equation}
where $\rho(t)$ is the time-dependent one-body density matrix and $h[\rho(t)]$ 
is the time-dependent mean field, which is a functional of the time-dependent 
density matrix. Although the TDHF equation is much simpler to solve numerically 
and was, historically, the first version of TDDFT applied to fission studies 
\cite{Negele1978,Umar2010,Simenel2014}, it has strong limitations: unless the 
calculation is initialized rather close to the scission point, or with an 
artificial boost in energy, the time-evolution of fissile nuclei such as 
$^{240}$Pu does not lead to fission \cite{Goddard2015,Goddard2016}. In order 
to achieve actual scission, pairing correlations play an essential role 
\cite{Scamps2015a,Bulgac2016}. For this reason, it is preferable to use either 
the TDHF+BCS theory \cite{Ebata2010,Scamps2012} or the fully-fledged TDHFB 
formalism. The TDHFB equation is formally similar to (\ref{eq:tdhf}) as it 
reads
\begin{equation}
i\hbar \frac{\partial\mathcal{R}}{\partial t} = \big[ \mathcal{H}(t), \mathcal{R}(t) \big],
\label{eq:tdhfb}
\end{equation}
where $\mathcal{R}(t)$ is the time-dependent generalized density matrix and 
$\mathcal{H}(t)$ the time-dependent HFB matrix; cf. (\ref{eq:HFB_matrices}). 
When setting the initial condition of the TDHFB equation near the saddle point 
of the potential energy surface (see Figure \ref{fig:physics}), the density of 
actinide nuclei will evolve toward two separated fragments as represented in 
Fig.\ref{fig:tddft}. In a sense, TDDFT generates fission events that are 
somewhat similar to Langevin trajectories.

One of the key properties of TDDFT is that the total energy as computed from 
(\ref{eq:HFB_energy}) (only, with time-dependent densities) is a constant of 
motion, $E(t) = E$. This means that if at $t=0$ the system is initialized in a 
static configuration $\gras{q}$ of energy $E_0 = E(\gras{q}(t=0))$, 
corresponding to the top of the fission barrier, at later times it will have 
acquired an excitation energy $E^* = E_0 -E(\gras{q}(t))$ since the potential 
energy tends to decrease with deformation. This property is the direct 
consequence of the fact that TDDFT simulates the diabatic evolution of the 
nucleus. It is a major advantage of TDDFT: when the fragments are formed at 
scission, the sharing of the prescission excitation energy will be done 
``automatically'' based on nuclear forces and quantum many-body effects rather 
than empirical formulas based on physics intuition.

In spite of its many advantages, TDDFT also has limitations. By construction, 
it is designed to simulate the time-evolution of the expectation value 
$\braket{\hat{A}}$ of one-body observables (TDHF) or two-body observables 
(TDHFB, with some simplifications) \cite{Blaizot1985,Balian1985}. It is not, 
however, well adapted to describe the fluctuations of these observables, that 
is, quantities $\sigma^{2}_{A} = \braket{\hat{A}^2} - \braket{\hat{A}}^2$; see 
discussions in \cite{Balian1984,Blaizot1985,Balian1992}. This has practical 
consequences: when solving the TDHFB equation with a broad range of initial 
conditions at the beginning of the main fission valley in actinides, the 
resulting trajectories tend to bundle and converge to very similar solutions 
\cite{Bulgac2019a}. In other words, TDHFB is very good at simulating the most 
likely fission but does a poor job at exploring less likely fragmentations. 
Different solutions have been proposed, including the stochastic mean field 
\cite{Ayik2008,Lacroix2014,Tanimura2017} or adding empirical terms to simulate 
fluctuations and dissipation to the TDHFB equation \cite{Bulgac2019}. Another 
route, which we will discuss in the next section, consists in adopting a 
theoretical framework where collective correlations are explicitly included.

\subsection{Collective Models}

Time-dependent density functional does not assume the existence of collective 
variables. Although this may seems like an advantage of the approach since it 
removes some arbitrariness in defining what these variables should be, it turns 
out to be also a handicap to describe very collective phenomena like fission. 
The adiabatic time-dependent Hartree-Fock-Bogoliubov (ATDHFB) and the generator 
coordinate method with the Gaussian overlap approximation (GCM+GOA) incorporate 
collective variables $\gras{q}$ from the get-go, which allows them to probe 
collective correlations that may be out of reach of standard implementations of 
TDDFT.

\subsubsection{Adiabatic Models}

Early on, when solving the TDDFT equation was computationally intractable, 
there were many attempts to extract from the TDDFT equation a collective model 
that could be adapted to describing large-amplitude collective motion. The 
adiabatic time-dependent Hartree-Fock (ATDHF) and its extension to include 
pairing correlations (ATDHFB) provided many insights into such an approach 
\cite{Krieger1974,Brink1976,Villars1977,Baranger1978}. Here, we only outline 
the main concepts of these theories and focus on ATDHFB; additional details can 
be found in the aforementioned references as well as the more recent reviews 
\cite{Nakatsukasa2016a,Schunck2019}. In TDHFB, the system explores a large 
variational space during its evolution; in ATDHFB, one seeks to force the 
system to follow specific trajectories within a collective subspace $\Sigma$ of 
the entire Hilbert space which, ideally, is decoupled from the rest of the 
space. In the seminal paper of Baranger and Veneroni on ATDHF 
\cite{Baranger1978}, this was achieved by introducing a unitary transform of 
the TDHF density $\rho(t)$. In the case of ATDHFB, this transformation reads: 
$\mathcal{R}(t)\rightarrow e^{i\hat{\chi}(t)}\mathcal{R}(t)e^{-i\hat{\chi}(t)}$ 
where $\hat{\chi}(t)$ is a one-body operator that is supposed to be small. The 
transformed density is expanded up to second order in $\hat{\chi}$ together 
with the time-dependent HFB matrix, and the results are inserted into the TDHFB 
equation. Separating contributions that are time-odd from those that are 
time-even, we obtain the two coupled equations
\begin{subeqnarray}
i\hbar\dot{\mathcal{R}}^{(0)} & = &
\big[ \mathcal{H}^{(0)}, \mathcal{R}^{(1)} \big] 
+ 
\big[ \mathcal{H}^{(1)}, \mathcal{R}^{(0)} \big] , 
\slabel{eq:atdhfb_eq_1}
\\
i\hbar\dot{\mathcal{R}}^{(1)} & = &
\big[ \mathcal{H}^{(0)}, \mathcal{R}^{(0)} \big] 
+ 
\big[ \mathcal{H}^{(0)}, \mathcal{R}^{(2)} \big] 
+ 
\big[ \mathcal{H}^{(1)}, \mathcal{R}^{(1)} \big] 
+ 
\big[ \mathcal{H}^{(2)}, \mathcal{R}^{(0)} \big] .
\slabel{eq:atdhfb_eq_2}
\end{subeqnarray}
These are the ATDHFB equations. The total collective energy of the system can 
be expressed as a function of the various components of the generalized density
as
\begin{equation}
\mathcal{E} = \mathcal{V} 
+
\frac{1}{2}\text{Tr} \left( \mathcal{H}^{(0)}\mathcal{S}^{(2)} \right)
+ 
\frac{1}{2}\text{Tr} \left( \mathcal{K}^{(1)}\mathcal{S}^{(1)} \right) ,
\label{eq:atdhfb_energy}
\end{equation}
where $\mathcal{V}$, the collective potential energy, is in fact the HFB energy 
computed with $\mathcal{R}^{(0)}$. The kinetic energy term is the sum of the 
last two terms, which involve the matrices 
\begin{equation}
\mathcal{S}^{(n)} = \left( \begin{array}{cc}
\rho^{(n)} & \kappa^{(n)} \\
-\kappa^{(n)*} & -\rho^{(n)}
\end{array}\right), 
\quad
\mathcal{K}^{(n)} = \left( \begin{array}{cc}
\Gamma^{(n)} & \Delta^{(n)} \\
-\Delta^{(n)*} & -\Gamma^{(n)} .
\end{array}\right) .
\end{equation}
These are the $n^{\rm th}$ order term of the expansion of the TDHFB generalized 
density ($\mathcal{S}^{(n)}$) and HFB matrix ($\mathcal{K}^{(n)}$). Note that 
the diagonal elements of the $0^{\rm th}$ order term are slightly different 
since they look formally more like (\ref{eq:HFB_matrices}). In these 
expressions, all terms are time-dependent. 

In principle, the ATDHFB equations should be solved self-consistently. In 
practice, however, the ATDHFB theory is mostly used to extract an expression 
for the collective inertia tensor. To this end, several approximations are 
invoked. The most important one consists in reducing the collective motion to 
a small set of collective variables $\gras{q}$ and assuming that 
\begin{equation}
\dot{\mathcal{R}}_0 = \sum_{\alpha} \frac{\partial\mathcal{R}_0}{\partial q_{\alpha}} \dot{q}_{\alpha} .
\end{equation}
In other words, the variations of the generalized density are constrained 
within the manifold spanned by the vector $\gras{q}$. It is then possible to 
rewrite the total collective energy (\ref{eq:atdhfb_energy}) as $\mathcal{E} = 
\mathcal{V} + \mathcal{K}$ where
\begin{equation}
\mathcal{K} = \frac{1}{2} \sum_{\alpha\beta} M_{\alpha\beta} \dot{q}_{\alpha}\dot{q}_{\beta}
\end{equation}
is the total kinetic energy and $\tens{M}\equiv M_{\alpha\beta}$ is the 
collective mass tensor given by
\begin{equation}
M_{\alpha\beta} = 
\frac{\partial\mathcal{R}^{(0)}}{\partial q_{\alpha}}
\mathcal{M}^{-1}
\frac{\partial\mathcal{R}^{(0)}}{\partial q_{\beta}} .
\label{eq:M}
\end{equation}
Without additional approximations, the expression for $\tens{M}$ is not trivial 
as it involves the inverse $\mathcal{M}^{-1}$ of the QRPA matrix. The very 
first exact calculation of $\tens{M}$ has thus been reported only very recently 
\cite{Washiyama2021}. Instead it is customary in most applications to fission 
to adopt the cranking approximation, where the QRPA matrix is reduced to its 
diagonal form (which can easily be inverted) and the derivatives in 
(\ref{eq:M}) are evaluated numerically. Analytical expressions requiring only 
HFB solutions at a single point $\gras{q}$ in the collective space can be 
obtained by using linear response theory to express all derivatives in terms 
of the moments of the constraint operators $\hat{Q}_{\alpha}$
\begin{equation}
M_{\alpha\beta}^{(-n)} = 
\sum_{\mu<\nu} 
\frac{ \braket{\Phi(\gras{q}) | \hat{Q}_{\alpha}^{\dagger} | \mu\nu}\braket{\mu\nu |\hat{Q}_{\beta}|\Phi(\gras{q}) }}{(E_{\mu} + E_{\nu})^n},
\label{eq:moments}
\end{equation}
where $\ket{\mu\nu}$ is a two-quasiparticle excitation of the state 
$\ket{\Phi(\gras{q})}$ and $E_{\mu}$ and $E_{\nu}$ are the energies of the q.p. 
$\mu$ and $\nu$. In that case, the collective mass tensor becomes
\begin{equation}
\tens{M} = \hbar^2 \big[ M^{(-1)} \big]^{-1} M^{(-3)} \big[ M^{(-1)} \big]^{-1}
\label{eq:atdhfb_inertia}
\end{equation}
with the moments given by (\ref{eq:moments}). The perturbative cranking 
approximation is numerically convenient but seems to have a large effect on the 
determination of spontaneous fission paths as reported in \cite{Sadhukhan2013,
Giuliani2018a}. 

The ATDHFB formula is probably the most commonly used in applications to 
fission, whether to compute spontaneous fission half lives or induced fission 
fragment distributions. Although it is known to give the correct collective 
mass at least in the case of translations, as applied in practical calculations 
the ATDHFB theory does not enforce a complete decoupling between collective and 
non-collective motion: doing so would not only require solving the ATDHFB 
equation (\ref{eq:atdhfb_eq_1})-(\ref{eq:atdhfb_eq_2}) self-consistently rather 
than on a precomputed potential energy surface, it would also face conceptual 
problems caused by symmetry breaking; see discussion in Chapter 6 of 
\cite{Schunck2019}. The adiabatic self-consistent collective (ASCC) model 
solves many of these difficulties \cite{Matsuo2000,Hinohara2007,Hinohara2008} 
but has not been applied yet on realistic simulations of fission. Another 
limitation of the ATDHFB framework is that it assumes that at each point 
$\gras{q}$, the system is in its lowest energy state. Extensions of the 
formalism to include thermal excitations and diabatic motion have been proposed 
but not tested in practice \cite{Reinhard1986}.

\subsubsection{Generator Coordinate Method}

The Hill-Wheeler-Griffin equation of the GCM is not solvable analytically and 
its physical content is not immediately visible. This motivated searches for 
suitable approximations. The Gaussian overlap approximation (GOA) has been the 
most popular \cite{Brink1968,Onishi1975}. It consists in assuming that the 
norm overlap is a generalized Gaussian function of $\gras{q}-\gras{q}'$ of the 
type 
\begin{equation}
\mathcal{N}(\gras{q},\gras{q}') 
\propto 
\exp\left[ -\frac{1}{2}\sum_{\alpha\beta}(q_{\alpha}-q'_{\beta})G_{\alpha\beta}(q_{\beta} - q'_{\beta})\right] ,
\label{eq:GOA1}
\end{equation}
where $\tens{G} \equiv G_{\alpha\beta}$ is called the metric tensor (in 
the most general case, it can depend on $\gras{q}$), and that the energy kernel 
can be written
\begin{equation}
\mathcal{H}(\gras{q},\gras{q}') = \mathcal{N}(\gras{q},\gras{q}')h(\gras{q},\gras{q}') ,
\label{eq:GOA2}
\end{equation}
where $h(\gras{q},\gras{q}')$ is a polynomial of degree 2 for the collective 
variables $\gras{q}$ and $\gras{q}'$, which is called the reduced Hamiltonian. 
The main advantage of the GOA is that it allows transforming the 
Hill-Wheeler-Griffin equation into a Schr\"odinger-like collective equation 
for the function $g(\gras{q})$ that is related to the weight function through
\begin{equation}
g(\gras{q}) = \int d\gras{q}' \, \mathcal{N}^{1/2}(\gras{q},\gras{q}') f(\gras{q}').
\label{eq:g_func}
\end{equation}
The function $g(\gras{q})$ can be interpreted as a probability to be at point 
$\gras{q}$ in the collective space. The GCM+GOA provides a general framework to 
compute the collective inertia tensor and various zero-point energy corrections 
associated with the collective motion \cite{Reinhard1987}. Indeed, after using 
the GCM ansatz (\ref{eq:GCM_ansatz}) for the many-body wave function 
$\ket{\Psi}$ and inserting the GOA approximations 
(\ref{eq:GOA1})-(\ref{eq:GOA2}), one can show (after some rather lengthy 
derivations presented in \cite{Ring2004,Krappe2012}) that the expectation value 
of the total energy $\braket{\Psi|\hat{H}|\Psi}$ becomes
\begin{equation}
\braket{\Psi|\hat{H}|\Psi}
=
\int d\gras{q} g^{*}(\gras{q}) \mathcal{H}_{\rm coll}(\gras{q})g(\gras{q}),
\end{equation}
where the collective Hamiltonian $\mathcal{H}_{\rm coll}(\gras{q})$ reads
\begin{equation}
\mathcal{H}_{\rm coll}(\gras{q})
=
-\frac{\hbar^2}{2} \sum_{\alpha\beta} 
\frac{\partial}{\partial q_{\alpha}}
B_{\alpha\beta}(\gras{q}) 
\frac{\partial}{\partial q_{\beta}}
+
\mathcal{V}(\gras{q}) .
\label{eq:GOA_H}
\end{equation}
In this expression, which corresponds to the special case of a constant metric 
$\tens{G}$, $\tens{B} = \tens{M}^{-1}$ is the collective inertia (the inverse 
of the collective mass) and $\mathcal{V}(\gras{q})$ is the collective 
potential. The collective potential is typically the sum of the HFB potential 
energy at point $\gras{q}$ and some zero-point energy corrections 
$\varepsilon_{\rm ZPE}(\gras{q})$. In a similar manner as for the ATDHFB 
theory, the calculation of the GCM+GOA collective mass and zero-point energy 
requires computing some derivatives with respect to the collective variables 
and dealing with the inverse of the QRPA matrix $\mathcal{M}$; see 
\cite{Schunck2016} for details. It is possible to adopt the same cranking and 
perturbative cranking approximation to simplify the calculations and make all 
quantities local to the point $\gras{q}$ in the coordinate space. In such cases, 
the collective mass and zero-point energy are given by
\begin{equation}
\tens{M} = 4 \tens{G} M^{(-1)} \tens{G}, 
\qquad
\varepsilon_{\rm ZPE}(\gras{q}) = \frac{1}{2} \tens{G}\tens{M}^{-1},
\end{equation}
with $M^{(-1)}$ the same moments defined above by (\ref{eq:moments}). In its 
common implementations, the GCM+GOA suffers from the same limitations as the 
ATDHFB theories: it only includes quasiparticle vacuua at point $\gras{q}$ and, 
therefore, cannot handle diabatic motion. This could be remedied either by 
enlarging the collective space to include quasiparticle excitations as proposed 
in \cite{Bernard2011}, or by developing a more general theory of quantum 
transport that combines the GCM with a finite-temperature formalism 
\cite{Dietrich2010}.

The GOA can also be applied to provide a time-dependent collective model. When 
the ansatz (\ref{eq:GCM_ansatz}) is inserted in the {\it time-dependent}
many-body Schr\"odinger equation, the resulting equation defines the 
time-dependent generator coordinate method (TDGCM) \cite{Verriere2020}. The GOA 
can then be applied to the TDGCM leading to the collective equation of motion
\begin{equation}
i \hbar\frac{\partial}{\partial t} g(\gras{q},t) 
= 
\left[ 
- \frac{\hbar^2}{2} \sum_{\alpha\beta} \frac{\partial }{\partial q_{\alpha}} B_{\alpha\beta}(\gras{q}) \frac{\partial}{\partial q_{\beta}} 
+
\mathcal{V}(\gras{q})
\right] g(\gras{q},t) .
\label{eq:evolution0}
\end{equation}
The right-hand side of this equation is exactly the same as (\ref{eq:GOA_H}), 
but the TDGCM+GOA equation now provides an evolution equation for the 
time-dependent probability of the system to be at a specific location 
$\gras{q}$ in the collective space at time $t$. Since the pioneering work of 
the CEA,DAM,DIF group at Bruy\`eres-le-Chatel \cite{Berger1984,Berger1986,
Goutte2004}, this technique has proven very successful to compute the 
distributions of fission fragments as we will see below.

\section{Selected Results}

\subsection{Spontaneous Fission}

As mentioned in the introduction, the description of spontaneous fission relies 
heavily on the concept of tunneling across multi-dimensional potential energy 
surfaces. Specifically, all realistic estimates of spontaneous fission 
half-lives published so far are based on the WKB formula to estimate the 
transmission coefficient through the barrier \cite{Landau1981}. The half-life 
is then given by the inverse of the transmission coefficient times the number 
of assaults to the barrier per unit time $\nu$ and reads
\begin{equation}
\tau_{\rm SF} = \frac{1}{\nu}\exp\left( 
\frac{2}{\hbar}\int_{a}^{b} ds\sqrt{2B(s)(\mathcal{V}(s) - E_0)}
\right) ,
\label{eq:t_sf}
\end{equation}
where $E_0$ is the ground-state energy and $\mathcal{V}$ the potential energy 
along the most likely fission path as parametrized by the curvilinear abscissa 
$s$. The quantity $B(s)$ is the collective inertia tensor that we have already 
mentioned several time. The integrand in (\ref{eq:t_sf}) is the classical 
action, and it is integrated between the inner turning point $s=a$ and the 
outer turning point $s=b$. The formula (\ref{eq:t_sf}) requires a collective 
space spanned by some variables $\gras{q}$, with a collective potential energy 
$\mathcal{V}(\gras{q})$ and a collective inertia tensor $\tens{B}(\gras{q})$. 
Since there are different models to compute these quantities (we mentioned the 
ATDHFB and GCM+GOA models) there will be a strong dependency of the calculated 
spontaneous fission half-lives on the model inputs -- especially so since both 
the potential and inertia enter as an exponent. For example, systematic 
calculations show very large variations of $\tau_{\rm SF}$ depending on: (i) 
the model used to compute the inertia (ATDHFB or GCM) and the accompanying 
approximations (perturbative versus non-perturbative cranking) 
\cite{Sadhukhan2013,Giuliani2018} (ii) the model to compute the collective 
energy $\mathcal{V}$ \cite{Bernard2019} (iii) the type and number of collective 
variables, especially the role of pairing correlations \cite{Sadhukhan2014,
Giuliani2014,Zhao2016}, (iv) the prescription for the ground-state energy 
\cite{Staszczak2009,Giuliani2018}.

\begin{figure}[!ht]
\centering
\includegraphics[width=1\textwidth]{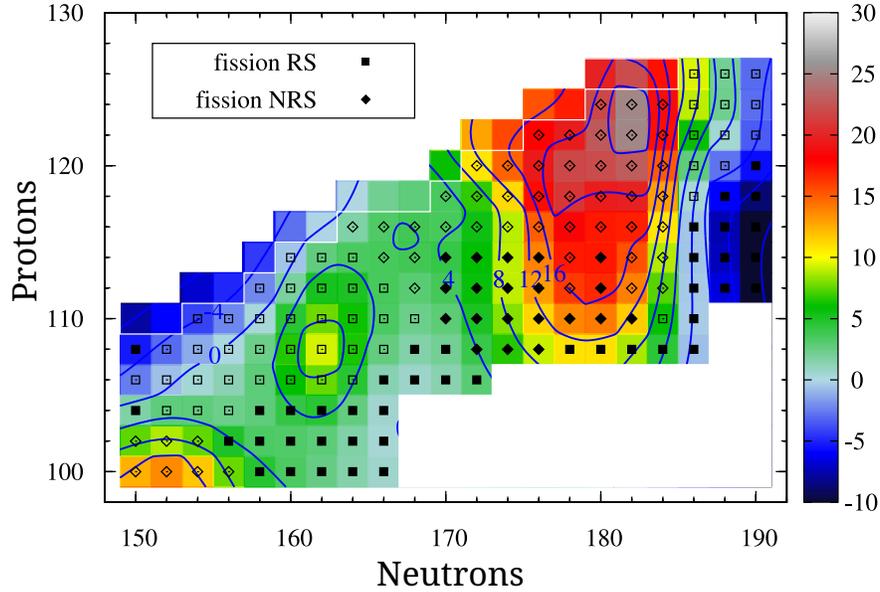}
\caption{Logarithm of the spontaneous fission half-lives of heavy and 
superheavy even-even nuclei with the D1S Gogny functional. Nuclides where
fission is the dominant decay mode are represented by filled symbols, those 
were $\alpha$ decay is dominant by open symbols. Squares indicate 
reflection-symmetric fission while diamonds indicate reflection-asymmetric 
fission.
Figures reproduced with permission from \cite{Baran2015} courtesy of 
Warda; copyright 2015 by Elsevier.
}
\label{fig:t_sf}
\end{figure}

In large-scale simulations, the spontaneous fission half-life is not computed 
from the most likely fission path but from the least-energy path. Finding the 
most likely fission path requires computing a $N$-dimensional collective space 
between the ground-state and the outer turning line and then selecting out of 
all possible trajectories the one that minimizes the classical action. The 
computational cost is substantial and for this reason, this method has not been 
applied to systematic calculations. Instead, such calculations are based on 
computing an effective one-dimensional fission path, typically parametrized by 
the axial quadrupole moment, and assume that such a path is a good 
approximation of the most likely fission path. Figure \ref{fig:t_sf} shows a 
representative example such a global calculation of $\tau_{\rm SF}$ for 
even-even heavy and superheavy elements in the case of the Gogny energy 
functional \cite{Baran2015}; similar systematics are also available for the 
Skyrme functional \cite{Staszczak2013}. Although 
agreement with experimental data (where available) may not always be better 
than a few orders of magnitude, one should keep in mind that, because of the 
exponential factor in (\ref{eq:t_sf}), half-lives are unforgiving tests of 
nuclear models. One of the advantages of microscopic methods is that they 
provide a consistent, rigorous and improvable framework to evaluate 
$\tau_{\rm SF}$. As computational capabilities increase, it should soon become 
possible to extract genuine least-action trajectories from multi-dimensional 
potential energy surfaces for an entire section of the nuclear chart (instead 
of a few isotopes) and compute $\tau_{\rm SF}$ with exact ATDHFB inertia and 
energy functionals properly calibrated to nuclear deformation properties. Even 
if the WKB approximation proves insufficient, there are already several other 
avenues worth exploring such as functional integral methods \cite{Levit1980,
Levit1980a,Levit1980b,Puddu1987,Levit2021}, instantons \cite{skalski2008,
brodzinski2020}, time-dependent Schr\"odinger with a complex absorbing 
potential \cite{Scamps2015}, configuration-interaction approach 
\cite{Hagino2020,Hagino2020a}. 

While half-lives have been the focus of most studies of spontaneous fission, 
they are not the only observables in this process. Fission fragment 
distributions are also important, especially for applications to, e.g., 
nucleosynthesis modeling where fission fragments may re-enter the set of 
reactions forming heavy elements \cite{Vassh2020,Cowan2021}. Such distributions 
also probe the influence of the entrance channel and/or the dependence of 
fission fragment distributions on the excitation energy of the nucleus. For 
example, neutron reactions on $^{239}$Pu lead to the fission of $^{240}$Pu with 
prescission energies of the order of 20-25 MeV, while the spontaneous fission 
of that same $^{240}$Pu happens at much lower prescission energies, of the 
order of 15-20 MeV. The determination of spontaneous fission fragment 
distributions can be decomposed in a two-step process: (i) the calculation of 
the tunneling probabilities to go from the ground state to the outer turning 
line and (ii) the evolution of the system from the outer turning line to 
scission. In \cite{Sadhukhan2016}, the authors propose to use a 
semi-microscopic method based on combining tunneling probabilities computed 
with the WKB formula and microscopic potential energy and collective inertia, 
and a semi-classical Langevin evolution from outer turning line to scission. 

\begin{figure}[!ht]
\centering
\includegraphics[width=0.45\textwidth]{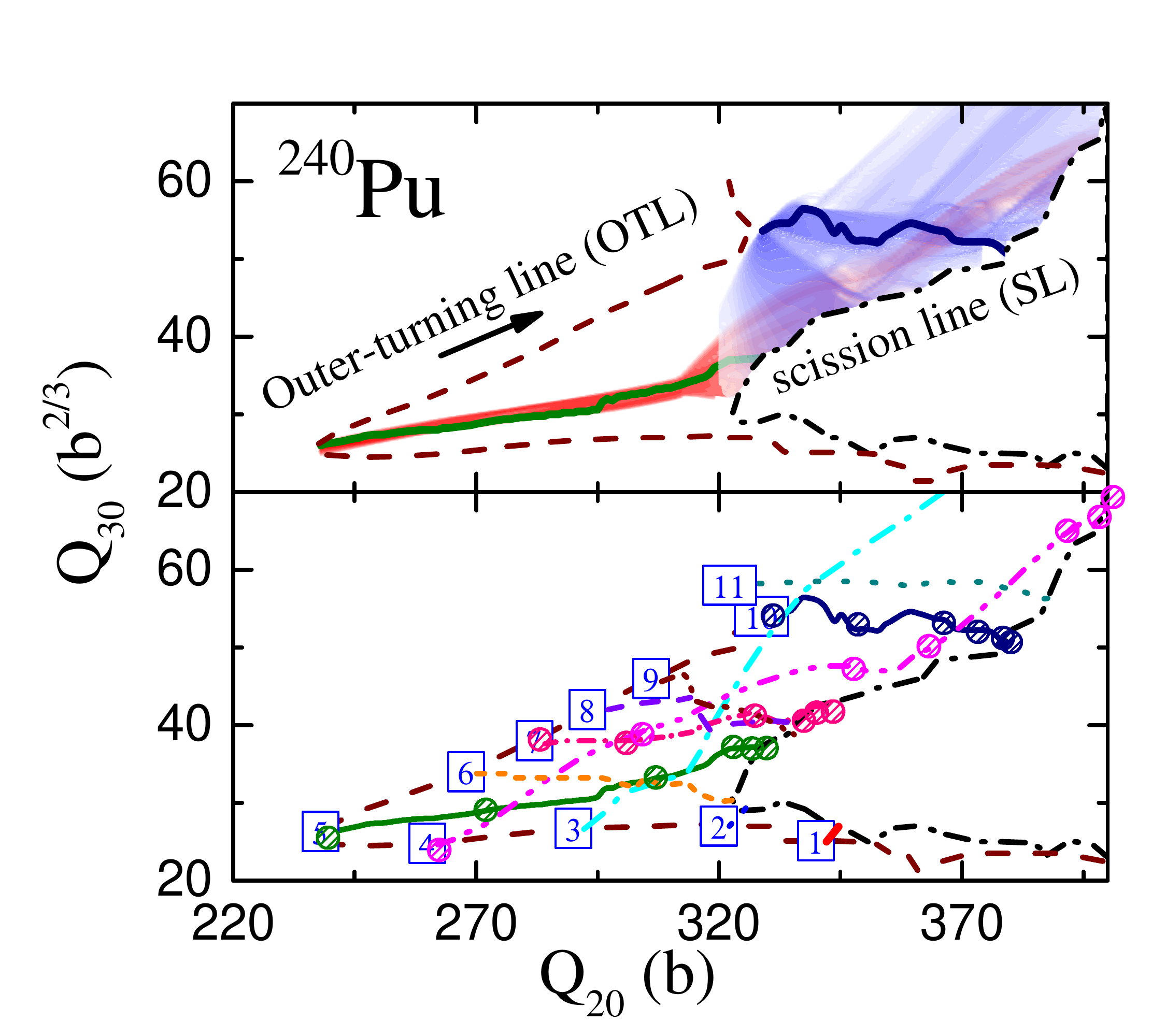}
\includegraphics[width=0.54\textwidth]{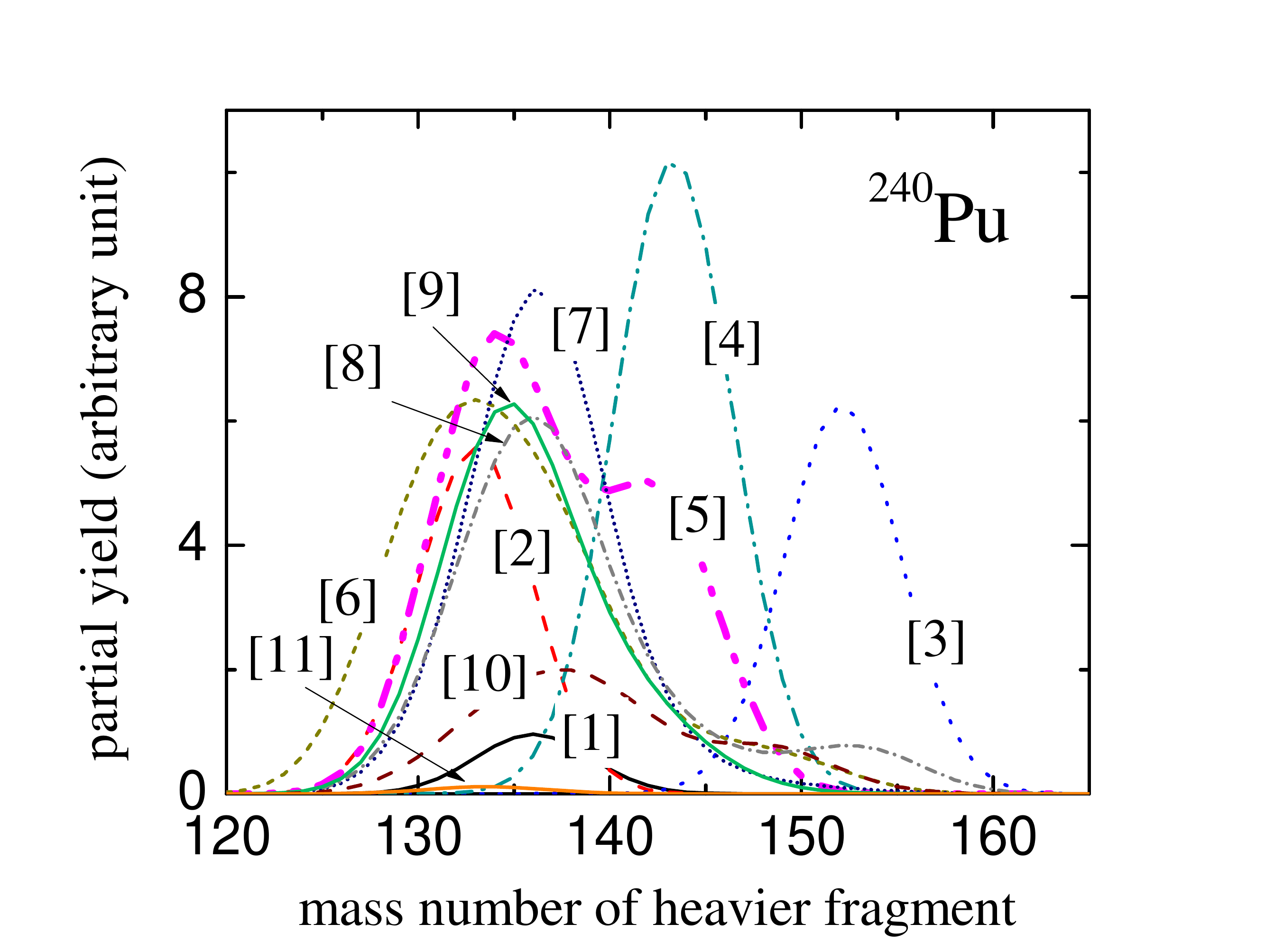}
\caption{Left: Effective fission paths from the outer-turning line (brown 
dashed line) to scission (black dashed-dotted line) in $^{240}$Pu. 
Right: Contribution to the fission fragment distribution of each effective 
fission path
Figures reproduced with permission from \cite{Sadhukhan2017} courtesy of 
Sadhukhan; copyright 2017 by APS.
}
\label{fig:EFP}
\end{figure}

A subsequent study showed that non-classical Langevin trajectories (where the 
energy may increase because of fluctuations) played an essential role in 
setting the tails of the fission fragment distributions, which correspond to 
very asymmetric fission. This is illustrated in Fig.\ref{fig:EFP}. The left top 
panel shows the outer turning line and scission line in a two-dimensional 
potential energy surface in $^{240}$Pu. The bottom left panel shows eleven 
different families of Langevin trajectories that originate from different 
initial points at the outer turning line. Most of the trajectories follow the 
slope of the potential energy surface, but a few (most notably numbers 3 and 
4), are non-classical, in the sense that the energy along the trajectory may 
increase. This behavior is the direct consequence of the large 
fluctuation-dissipation term in the Langevin equation. What the right panel of 
the figure shows is that these non-classical trajectories give contributions to 
the very asymmetric part of the fission fragment distributions. 

\subsection{Characterization of Fission Fragments}

\subsubsection{Number of Particles}

\begin{figure}[!ht]
\centering
\includegraphics[width=0.8\textwidth]{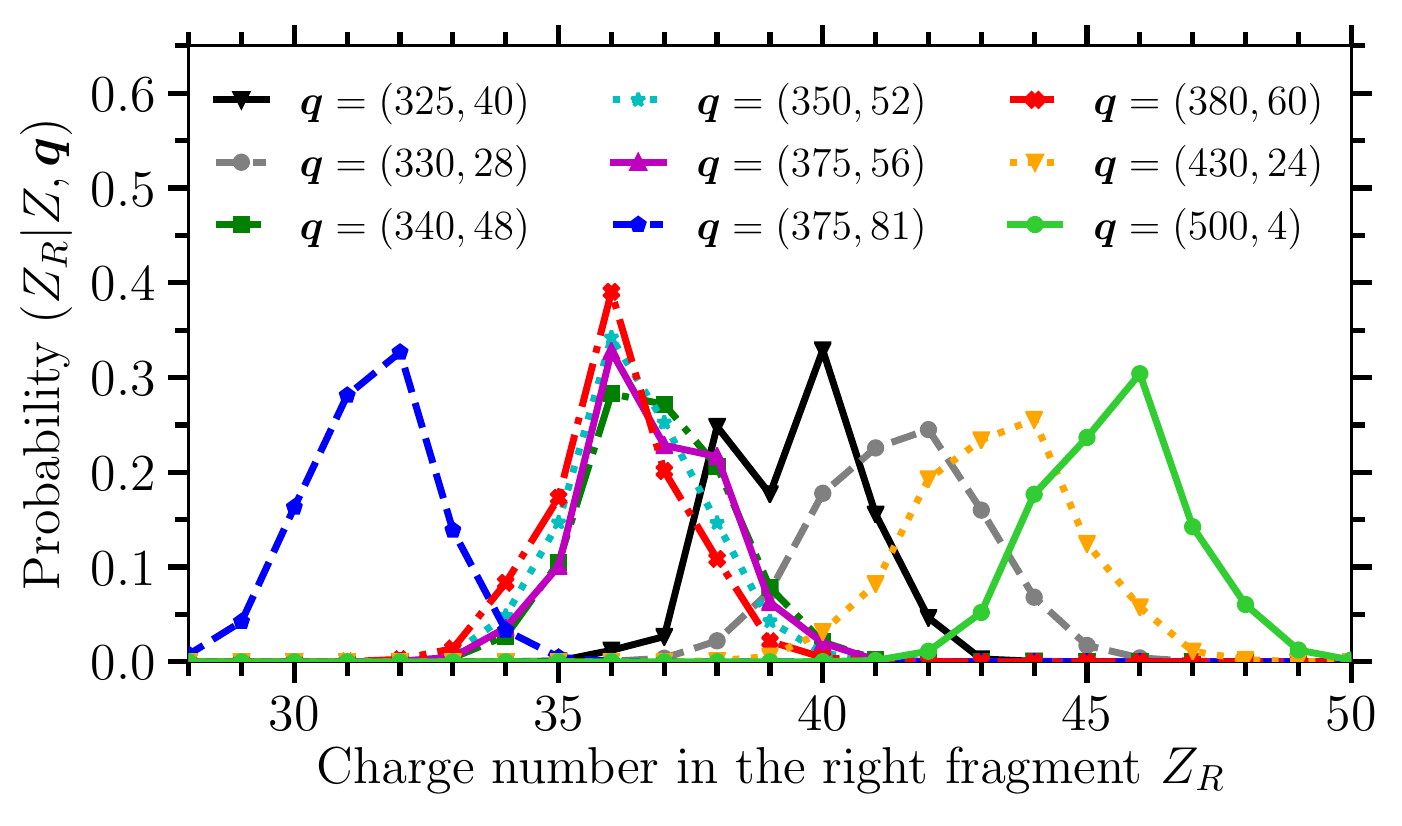}
\caption{Probability distribution $\mathbb{P}(Z_{\rm R}|  Z, \gras{q})$ of 
Eq.~(\ref{eq:probaZ}) that a {\it single} scission configuration $\gras{q}$ 
contains $Z_{\rm R}$ protons. Each curve corresponds to different scission 
configuration for the thermal fission of $^{235}$U(n,f).
Figures reproduced with permission from \cite{Verriere2021a} courtesy of 
Verriere; copyright 2021 by APS.
}
\label{fig:part_fragments}
\end{figure}

For a long time, the number of particles in fission fragments was obtained in a 
semi-classical way by integrating the density of particles (proton, neutron or 
total density) in the regions of space associated with each fragment. If we 
note $z_{\rm N}$ the position along the $z$-axis where the density is the 
lowest between the two prefragments at scission, then the average number of 
particles in the right fragment would be
\begin{equation}
\bar{X}_{\rm R} = \int_{-\infty}^{+\infty} dx\int_{-\infty}^{+\infty} dy \int_{z_{\rm N}}^{+\infty} dz\, \rho_{X}(\gras{r})
\label{eq:ZR_avg}
\end{equation}
where $\rho_{X}(\gras{r})$ is the density of particles $X=n,p$. As emphasized 
by the notation $\bar{\dots}$, the quantity thus computed is an average value. 
Among other things, it may not be an integer number. A much better prescription 
based on particle number projection techniques was proposed originally in 
\cite{Simenel2011}. The basic idea is to use the formalism very briefly 
outlined above with the following modification: the particle number operator is 
replaced by 
\begin{equation}
\hat{X}_{\rm R} = \sum_{\sigma} \int d^{3}\gras{r}\,
H(z-z_{\rm N}) c^{\dagger}(\gras{r},\sigma)c(\gras{r},\sigma)
\label{eq:ZR_op}
\end{equation}
where $H(z-z_{\rm N})$ is the Heaviside function \cite{Abramowitz1964} and 
$c^{\dagger}(\gras{r},\sigma)$ creates a particle at point $\gras{r}$ with spin 
projection $\sigma$. By construction, the operator (\ref{eq:ZR_op}) counts the 
number of particles in the right fragment only, and it is easy to check that  
its expectation value is $\braket{\hat{X}_{\rm R}} = \bar{X}_{\rm R}$ as 
defined in (\ref{eq:ZR_avg}). From the definition (\ref{eq:proj}) specialized 
to the case (\ref{eq:PNP}) with the operator (\ref{eq:ZR_op}), we can then 
introduce the projection operator $\hat{P}^{(\rm R)}_{\rm X}(X_{\rm R})$ that 
extracts the probability that a wave function contains $X_{\rm R}$ particles in 
the right fragment. This allows us to define the probability to find 
$X_{\rm R}$ particles at point $\gras{q}$ in the collective space, given a 
total number of particles $X$ in the compound nucleus
\begin{equation}
\mathbb{P}(X_{\rm R}|  X, \gras{q}) 
=
\frac{\braket{\Phi(\gras{q}) | \hat{P}^{(\rm R)}_{\rm X}(X_{\rm R}) \hat{P}_{\rm X}(X) | \Phi(\gras{q})}}{\braket{\Phi(\gras{q}) | \hat{P}_{\rm X}(X) | \Phi(\gras{q})}}
\label{eq:probaZ}
\end{equation}
An example of this probability is represented in Fig.~\ref{fig:part_fragments}
for different scission configurations of the compound nucleus $^{236}$U formed 
in the thermal reaction $^{235}$U(n,f). One can notice the case of the 
$(325,40)$ configuration (the numbers represent, respectively, the average 
quadrupole moment $\braket{\hat{Q}_{20}}$  [b] and average octupole moment 
$\braket{\hat{Q}_{30}}$ [b$^{3/2}$] of the configuration): the probability 
shows a clear odd-even staggering. This effect has been observed experimentally 
in the charge distributions \cite{Ehrenberg1972,Amiel1975,Mariolopoulos1981,
Bocquet1989} and could only be reproduced theoretically by introducing 
{\it ad hoc} parameters. Results such as the ones shown in 
Fig.~\ref{fig:part_fragments} suggest that PNP may be the key ingredient needed 
to truly predict the odd-even staggering of charge distributions.

The first application of particle number projection in the fragments was done 
in \cite{Scamps2015a}. The formalism was presented in more details in 
\cite{Verriere2019} and was applied for the first time in the calculation of 
primary fission fragment charge, mass and isotopic distributions in 
\cite{Verriere2021a}. This technique has several advantages. First, it can be 
applied to phenomenological models based, e.g., on the macroscopic-microscopic 
model. The only ingredient needed is a set of single-particle levels associated 
with each configuration. Second, the method guarantees that the fragments will 
have an integer number of particles -- as nature tells us. Finally, by 
substituting at each scission point a single pair of average numbers 
$(\bar{Z}_{\rm R}, \bar{N}_{\rm R})$ by a two-dimensional probability 
distribution function $\mathbb{P}(Z_{\rm R},N_{\rm R}|  Z,N,\gras{q})$, which 
is equal to $\mathbb{P}(Z_{\rm R}| Z, \gras{q}) \times 
\mathbb{P}(N_{\rm R}| N, \gras{q})$ if calculations do not account for isospin 
mixing, the method produces much more realistic estimates of which fragments 
are populated. We will show later how this technique can be applied to compute 
isotopic yields $Y(Z,A)$ of neutron-induced fission reactions.

\subsubsection{Deformations}

As mentioned during the presentation of the formalism of the HFB theory, the 
deformation of the nuclear shape in EDF approaches is set by imposing 
constraints on the expectation value of suitable operators. This can be viewed 
as somewhat equivalent to macroscopic-microscopic methods, where the nuclear 
shape is parametrized with a set of parameters in such a way as to cover the 
range of shapes relevant in fission. The big difference between the two 
approaches, however, concerns the deformations that are {\it not} enforced 
specifically in the calculation. In macroscopic-microscopic models, two options 
are possible: if the shape parametrization has a finite number of parameters, 
as in most applications of fission, then any shape {\it not} represented by 
this parametrization can not be included in the calculation; if the 
parametrization has an infinite number of parameters, as in the multipole 
expansion of the nuclear radius, then all parameters not set explicitly are 
zero by default. In contrast, the variational principle at play in the EDF 
approach ensures that all multipole moment expectation values 
$\braket{\hat{Q}_{\lambda\mu}}$ that are not explicitly set by the user will 
automatically take whichever value minimizes the total energy. When computing 
potential energy surfaces, this property can turn into a curse as it is the 
reason why discontinuities appear in the surface \cite{Dubray2012}. However, 
the same property turns into a blessing when looking at the properties of the 
fragments at scission.

\begin{figure}[!ht]
\centering
\includegraphics[width=0.75\textwidth]{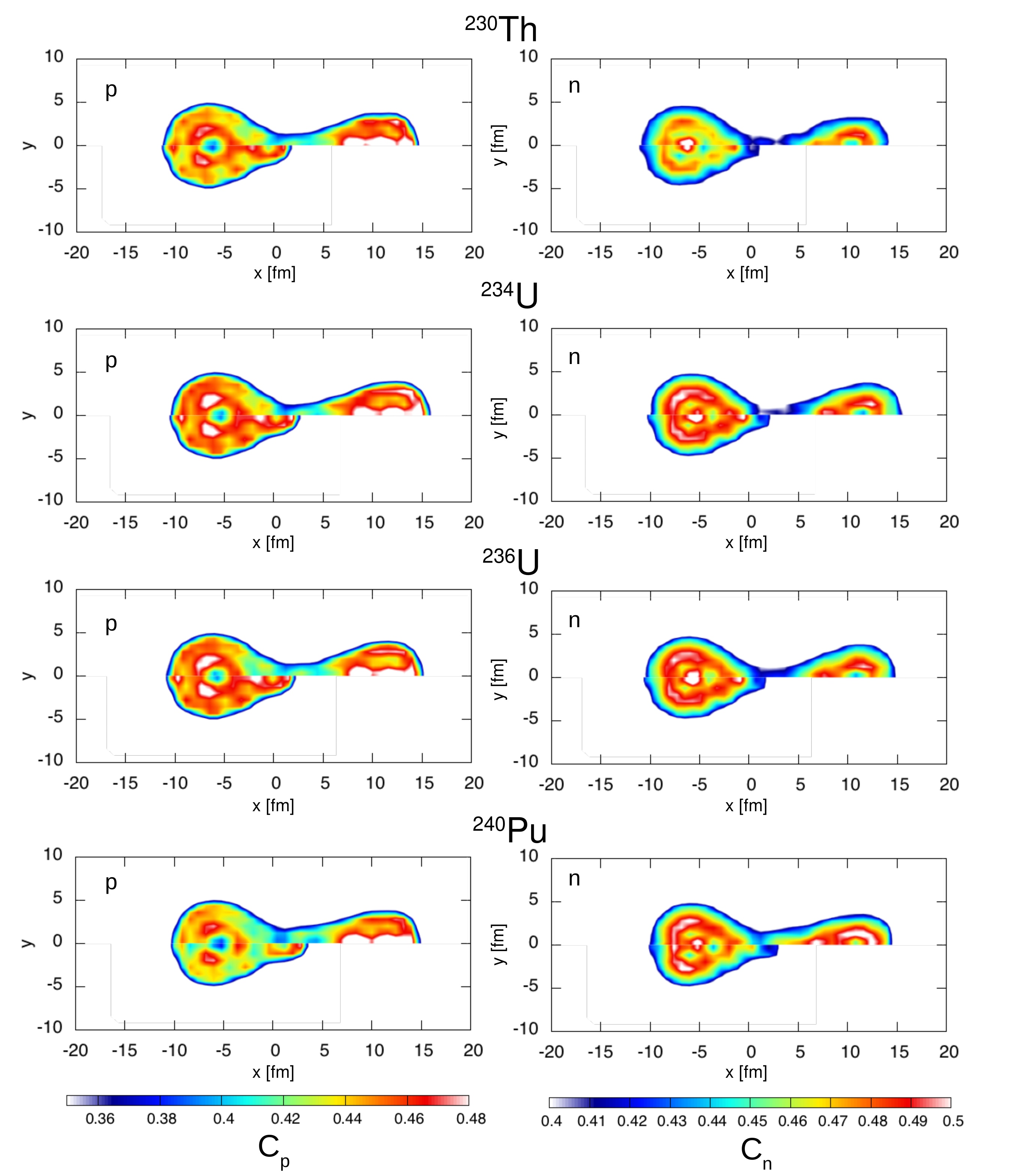}
\caption{Identification of the heavy pre-fragment in the asymmetric fission of 
major and minor actinides. The figure compares the localization functions of 
the fissioning nucleus at scission (upper half of the contour plot in each 
frame) with that of the octupole deformed $^{144}$Ba (lower-half of the contour 
plots) and highlights the role of octupole shell effects to determine the 
number of particles in the fragments.
Figures reproduced with permission from \cite{Scamps2018} 
courtesy of Scamps; copyright 2021 by Nature.
}
\label{fig:def_fragments}
\end{figure}

Figure \ref{fig:def_fragments} shows the localization functions of the left and 
right fragments of several actinide nuclei at scission \cite{Scamps2018}. The 
localization function is a visual indicator of not only the spatial profile of 
a density but also of it shell structure \cite{Reinhard2011}. The figure 
compares the proton $C_p$ and neutron $C_n$ localization functions of the 
fissioning nucleus and the one of the octupole-deformed fragment $^{144}$Ba, 
and shows that the heavy fragment in the most likely fragmentation tends to be 
octupole-deformed. The results reported in \cite{Scamps2018} explained for the 
first time why the peak of the fission fragment in actinides was not centered 
around $^{132}$Sn as would be expected if only {\it spherical} shell effects, 
which are maximum at $Z=50$ and $N=82$, were active in the fission fragments. 
Instead, they suggested that the peak of the distribution is slightly shifted 
toward heavier $Z$ and $N$ value because of the competition with {\it octupole} 
shell effects, which are maximum for $Z=56$ and $N=88$ \cite{Butler1996}. These 
results were only possible because in EDF calculations, the deformation of the 
fragments is not set beforehand but is an outcome of the variational 
calculation: fragments may be spherical, quadrupole-deformed or 
octupole-deformed depending on their number of protons and neutrons. This is a 
compelling illustration of the predictive power of microscopic methods.

\subsubsection{Spin Distributions}

The prediction of the spin distributions of fission fragment is another, quite 
recent success of microscopic methods based on the EDF approach. The large 
number of photons emitted during the deexcitation of fission fragments is 
experimental evidence that fragments have a rather broad spin distribution. 
Until recently, the functional form of the spin distributions was based on the 
statistical model and read \cite{Bloch1954}
\begin{equation}
p(J) \propto (2J+1)\exp \left[ -\frac{1}{2}\frac{(J+\tfrac{1}{2})^2}{\sigma^2} \right] ,
\label{eq:spin_distribution}
\end{equation}
where $\sigma$ is called the spin-cutoff parameter and is proportional to the 
expectation value $\braket{\hat{J}^{2}}$ of the total angular momentum in the 
system \cite{Madland1977,Bertsch2019}. In all deexcitation models of the 
nucleus, the parameter $\sigma$ is adjusted, together with other parameters 
pertaining to the decay, to match experimental data. Recent extensions of 
angular momentum projection in the fission fragments suggest that this 
parameter could be computed directly, or that (\ref{eq:spin_distribution}) 
itself could be extracted from the projection on good $J$ of scission 
configurations \cite{Bulgac2021,Marevic2021}.

The extraction of the probability distribution (\ref{eq:spin_distribution}) is 
analogous to the case of particle number. It is based on the angular momentum 
projection techniques of the MR-EDF theory presented earlier. Following 
\cite{Sekizawa2017}, one can define an angular momentum operator kernel acting 
only in the right fragment as
\begin{equation}
\gras{J}^{F}(\gras{r},\sigma) = H^{*}(z-z_N) \gras{J}(\gras{r},\sigma) H(z-z_N),
\label{eq:kernel}
\end{equation}
where $\gras{J}(\gras{r},\sigma) = \gras{L}(\gras{r}) + \gras{S}(\sigma)$ is 
the usual angular momentum operator acting on both spatial and spin coordinates 
and $H(z-z_N)$ is, as before, the Heaviside step function. Assuming that the 
$z$-axis of the fissioning nucleus intrinsic frame is the axis of elongation, 
the center of mass of each fragment is located at $\gras{r}_{\text{CM}}^{F} = 
(0, 0, z_{\text{CM}}^{F})$. Therefore, we must take $\gras{r} \rightarrow 
\gras{r} - \gras{r}_{\text{CM}}^{F}$ in (\ref{eq:kernel}) to determine the 
components of the angular momentum with respect to the center of mass of each 
fragment. To project on angular momentum in the fragments, the rotation 
operator of (\ref{eq:AMP}) must be defined with the components of the operator 
(\ref{eq:kernel}). The probability of finding spin $J$ in the fragment F at the 
scission configuration $\gras{q}$ is then
\begin{equation}
|c_{J}{\rm (F)}|^2 = 
\frac{2J+1}{16\pi^2}\int_{0}^{2\pi} d\alpha \int_{0}^{\pi} d\beta\sin\beta \int_{0}^{4\pi}d\gamma
\braket{\Phi(\gras{q}) | \hat{R}^{\rm (F)}(\Omega) | \Phi(\gras{q})} .
\label{eq:aJ}
\end{equation}
In the two applications of this technique reported so far \cite{Bulgac2021,
Marevic2021}, only axially-symmetric scission configurations were considered,
which reduced the full rotation operator to a rotation around the $y$-axis of
the reference frame: $\hat{R}^{\rm (F)}(\Omega) \rightarrow 
\hat{R}_{y}^{\rm (F)}(\beta)$. Just like in the case of particle number 
projection, the only ingredient to the method is a set of single- or 
quasi-particle states at scission making this technique applicable to 
macroscopic-microscopic models.

\begin{figure}[!ht]
\centering
\includegraphics[width=0.45\textwidth]{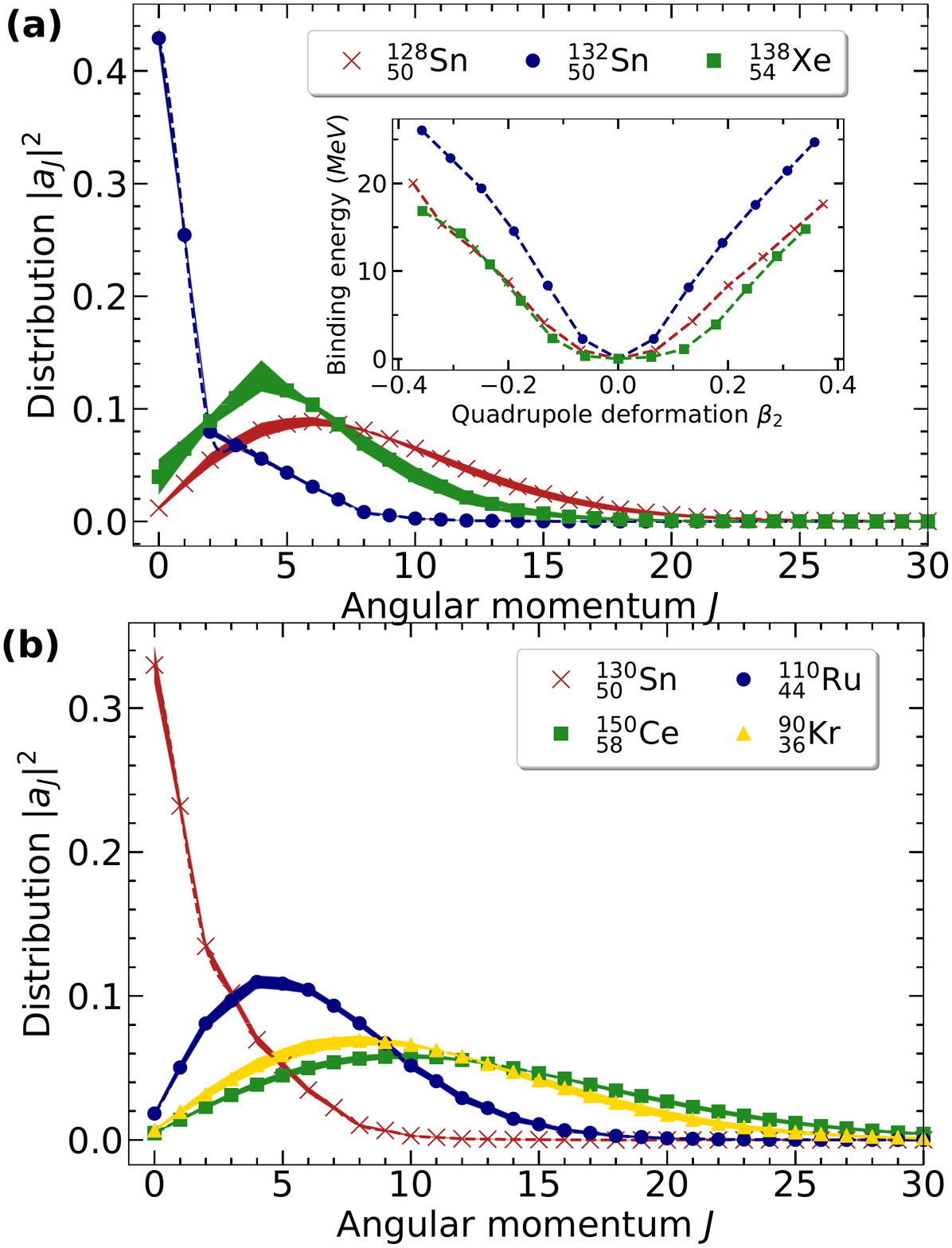}
\includegraphics[width=0.53\textwidth]{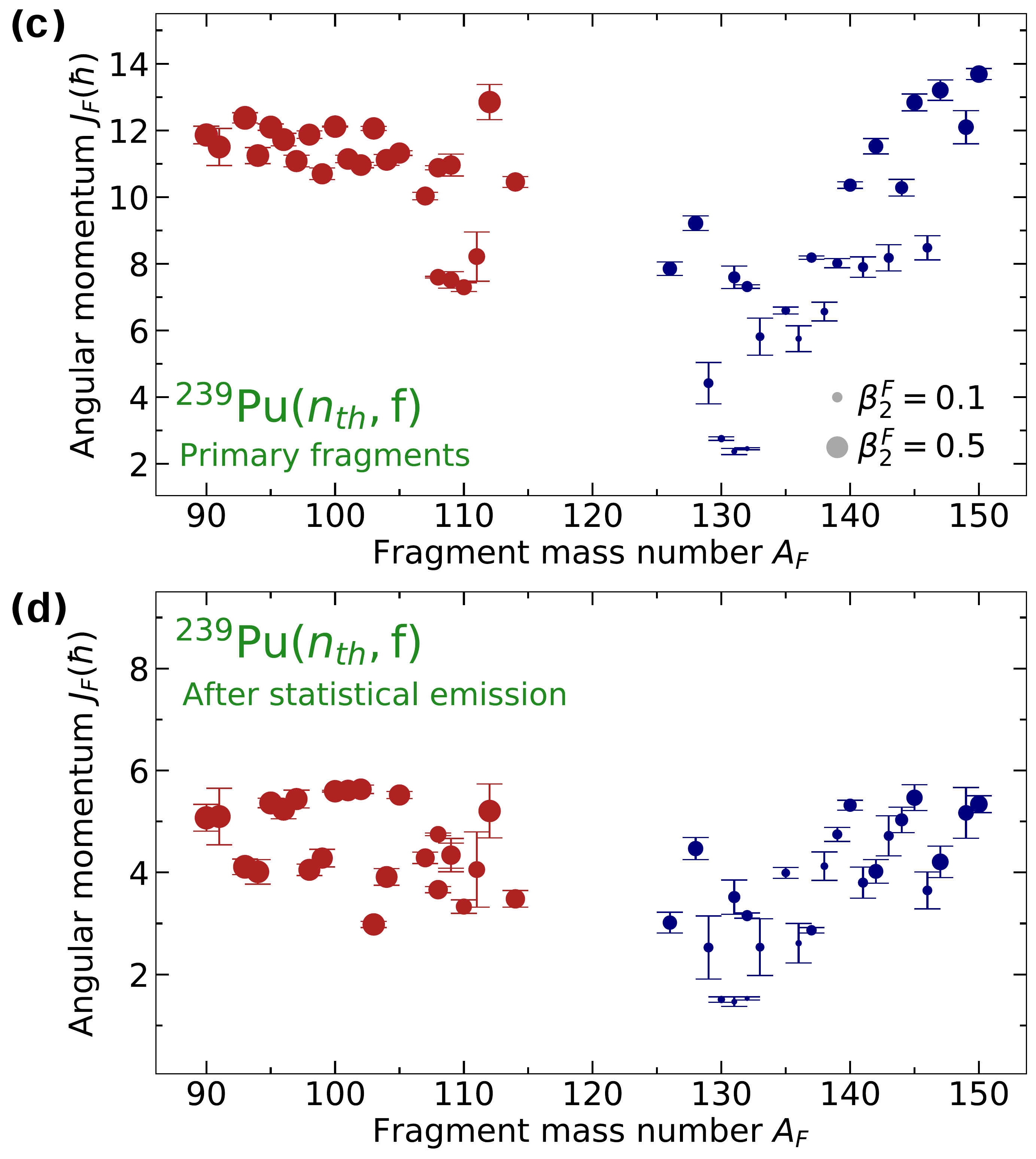}
\caption{Left column: spin distribution of several heavy fragments in the 
fission of $^{240}$Pu. Panel (a) highlights that nuclei that are spherical in 
their ground state may be substantially deformed at scission, which completely 
changes their spin distribution. Panel (b) illustrates that, on average, the 
spin of the light fragment is larger than the one of the heavy fragment. Right 
column: Average spin $\langle J\rangle(A)$ as a function of the mass of the 
fragment. Panel (c) shows the result before any emission of particle, while 
panel (d) shows the results after emission of neutrons and statistical photons.
Figures reproduced with permission from \cite{Marevic2021} courtesy of 
Marevic; copyright 2021 by APS.
}
\label{fig:spin_fragments}
\end{figure}

Figure \ref{fig:spin_fragments} shows an example of such calculations for the 
benchmark case of $^{240}$Pu. Panel (a) shows the spin distribution for three 
isotopes near the doubly-magic $^{132}$Sn nucleus. As recalled in the inset 
frame, all three isotopes are spherical in their ground state; at scission, 
however, their deformation can be different. While $^{132}$Sn is still almost 
spherical, which manifests itself by a sharply peaked distribution around 
$J=0$, the distribution of both $^{128}$Sn and $^{138}$Xe has a relatively 
large spread extending up to $J=20 \hbar$ for Xenon: this is indicative of a 
rather substantial deformation for that nucleus. Panel (b) highlights that the 
spin distribution of the fragments can vary substantially and, most 
importantly, that it does not depend only on the mass of the nucleus as often 
assumed in semi-empirical models of deexcitations \cite{Litaize2010,
Verbeke2018,Talou2021}: the curves for $^{150}$Ce and $^{90}$Kr are very 
similar, yet the latter is 50\% lighter than the former. Panels (c) and (d) on 
the right shows a systematics of the average spin of the fragments, that is, 
the quantity $\braket{J}$ defined through $\braket{J}(\braket{J}+1) = \sum_J 
|c_{J}{\rm (F)}|^2 J(J+1)$, as a function of the mass of that fragment. The 
size of each circle represents the quadrupole deformation $\beta_2$ of the 
fragment and the error bar the uncertainty of the calculation. Panel (c) shows 
$\braket{J}$ for the primary fragments, i.e., before any emission of particles 
take place; panel (d) shows the same quantity after neutrons and statistical 
photons have been emitted. The statistical deexcitation was simulated with the 
code FREYA \cite{Verbeke2018}. These results qualitatively reproduce (without 
adjustable parameters) recent experimental measurements \cite{Wilson2021}. 

The results reported in Fig.~\ref{fig:spin_fragments} were extracted from 
static calculations: angular momentum projection (in the fragments) was applied 
on HFB solutions corresponding to a large number of scission configurations 
\cite{Marevic2021}. While this allows extracting trends of $\braket{J}$ as a 
function of the mass or charge of the fragment, the downside of this approach 
is that the fragments are not fully separated and are ``cold'': all their 
excitation energy is in the form of deformation energy. In \cite{Bulgac2021}, 
the very same AMP technique was applied on the TDHFB solution for the most 
likely fission. As emphasized earlier, the total energy is a constant of motion 
in TDHFB simulations, hence the fragments have the right amount of excitation 
energy at scission. Interestingly, the spin distributions extracted from 
such ``hot'' fragments after full separation are very similar to the results 
obtained in ``cold'' fragments just before scission. This suggests that, as the 
shape of the fragments relax after the fragments separate, some deformation 
energy may be transferred into thermal excitation energy or, more rigorously, 
that the quantum fluctuations of $J$ induced by the large deformations are 
converted into statistical fluctuations \cite{Egido1988}.

\subsubsection{Excitation Energy}

The large number of emitted particles during the fission process is evidence 
that fission fragments are highly excited upon their formation. This simple 
fact can also be inferred from the energy balance of the fission reaction
\begin{equation}
\label{eq:energy_balance}
{\rm TXE} + {\rm TKE} + M(Z_{\rm H},N_{\rm H}) + M(Z_{\rm L},N_{\rm L})
= 
M(Z,N)  + S_{n}(Z,N) + E_n,
\end{equation}
where $M(Z,N)$ is the mass of the nucleus $(Z,N)$, $S_{n}(Z,N)$ the one-neutron 
separation energy in the nucleus $(Z,N)$, $E_n$ the energy of the incident 
neutron in the center of mass frame, TXE the total excitation energy available 
to both fragments, and TKE the total kinetic energy of the fragments. In the 
case of spontaneous fission, there is no incident neutron: $S_{n} = E_n = 0$. 
Similarly, for photofission (fission induced by photons), $S_{n} = 0$ and 
$E_n \equiv E_{\gamma}$. In the fission of actinides, the mass of nearly all 
fission fragments is known with excellent precision, as is the mass of 
fissioning nucleus and its one-neutron separation energy. Measurements of TKE 
indicate that this quantity is in the range 150--190 MeV. It is then easy 
to find that the total excitation energy, as indicated from relation 
(\ref{eq:energy_balance}), is of the order of 30--50 MeV. To take an example: 
in the thermal neutron-induced fission of $^{240}$Pu, we have $Z=94$ and 
$N=146$, $E_n = 0$ MeV and $S_n = 8.31$ MeV. Let us consider the pair of 
fragments $^{132}$Sn and $^{108}$Ru: the TKE is ${\rm TKE}(A=132) = 183.15$ MeV 
\cite{Tsuchiya2000}. Based on the values of all nuclear masses \cite{Wang2021}, 
we find ${\rm TXE} = 36.01$ MeV. This energy must be distributed between 
the fragments. In approaches based on identifying scission configurations from 
a potential energy surface, this problem is not easy to solve because, as 
recalled in the previous section, the fragments are cold by construction. For 
this reason, the energy is usually shared based on a statistical formula that 
depends on the deformation (at scission) and level density of each fragment 
\cite{Schmidt2011,Albertsson2020a}. In microscopic theories, this approach is 
complicated by the fact that estimates of quantities such as the energy, number 
and particles, etc., in the fragments are dependent on the degree of 
entanglement between them \cite{Younes2011,Schunck2014,Schunck2015b}.

Time-dependent DFT offers a very appealing solution to these problems. Recall 
that TDDFT equations simulate a complete fission event while, at the same time, 
conserving the total energy. If we wait long enough when running the 
simulation, we will reach the point where the two fragments are well separated 
with only the Coulomb force acting between them; see Fig.~\ref{fig:tddft}. The 
energy balance reads $E_0 = E_{\rm H}(t) + E_{\rm L}(t) + E_{\rm int}(t)$, 
where $E_0 = E(t=0)$ is the total, conserved energy of the system, 
$E_{\rm H,L}(t)$ is the time-dependent energy of the heavy/light fragment and 
$E_{\rm int}(t)$ is the interaction energy between them. At late times, or 
large distances between the two fragments, this interaction energy reduces to 
the total kinetic energy which is easily obtained from TDDFT evolution 
\cite{Simenel2014,Tanimura2015}. From that same evolution, we can extract the
density matrix $\rho_{\rm H,L}(t)$ and pairing tensor $\kappa_{\rm H,L}(t)$ of 
each fragment and therefore calculate their total energy directly. Subtracting 
the ground-state binding energy of each fragment gives the excitation energy 
$E^*{\rm H,L}$. Estimates of both TKE and the $E^*{\rm H,L}$ have been obtained 
for the case of neutron-induced fission of $^{239}$Pu and agree rather well 
with experimental measurements \cite{Bulgac2016,Bulgac2019a}. They also suggest 
that, at least for the most likely fission, the heavy fragment tends to have a 
lower internal temperature than the light one \cite{Bulgac2019a,Bulgac2020}. 
Note that the shape relaxation of the fragments and the large Coulomb 
interaction at scission complicate this picture since they lead to the 
excitation of giant resonances, which in turn imply that the energy of each 
fragment is not constant over time \cite{Simenel2014,Bulgac2020}. 

\subsection{Distribution of Fission Fragments}

The previous sections highlighted how microscopic methods based on the EDF 
approach can provide valuable insights on the properties of the fission 
fragments at the time of their formation, hence informing statistical models of 
deexcitation. Another piece of information that is crucial to model this decay 
is the relative population of each fission fragment. As mentioned in the 
introduction, this quantity is represented by the fission fragment 
distributions $Y(A)$ (mass distribution) $Y(Z)$ (charge distribution), and 
$Y(Z,A)$ (isotopic yields). Since the charge and mass of the fission fragments 
can be mapped onto the values of collective variables, predicting the relative 
population of the fragments is, to some extent, equivalent to predicting the 
population of specific points $\gras{q}$ of the collective space. The TDGCM is 
especially well suited to such a purpose, since the quantities 
$|g(\gras{q},t)|^2$ can precisely be interpreted as probabilities of occupation 
at point $\gras{q}$ and time $t$. 

The formalism to extract the distributions from a TDGCM+GOA evolution was 
presented in \cite{Goutte2004,Regnier2016b,Verriere2021a}, we only recall below 
the most important elements of the calculation. The (primary) yield of a 
fragment $(Z_{\rm f}, N_{\rm f})$ is the probability
\begin{equation}
Y(Z_{\rm f}, N_{\rm f}) = 100 \big[ 
\mathbb{P}(Z_{\rm f},N_{\rm f} | Z,N) 
+
\mathbb{P}(Z-Z_{\rm f},N-N_{\rm f} | Z,N) 
\big], 
\end{equation}
where $\mathbb{P}(Z_{\rm f},N_{\rm f} | Z,N)$ is the probability that the 
right fragment has $Z_{\rm f}$ protons and $N_{\rm f}$ neutrons and the 
compound nucleus has $Z$ protons and $N$ neutrons. We assume that this 
probability is given by
\begin{equation}
\mathbb{P}(Z_{\rm f},N_{\rm f} | Z,N) 
\propto
\int d\xi\,
F(\xi)
\mathbb{P}(Z_{\rm f},N_{\rm f} | Z,N, \gras{q}(\xi)) 
\end{equation}
where $\mathbb{P}(Z_{\rm f},N_{\rm f} | Z,N, \gras{q}(\xi))$ is the analogue of 
the probability (\ref{eq:probaZ}) and quantifies the probability that there is 
$Z_{\rm f}$ protons and $N_{\rm f}$ neutrons in the right fragment at the 
scission configuration $\gras{q}$. Here, we assume that the set of all scission 
configurations can be parametrized by the coordinate $\xi$ forming the scission 
``line'' $\mathcal{S} \equiv \{ \gras{q}(\xi) \}_{\xi}$. The quantity $F(\xi)$ 
is the time-integrated flux of the collective wave packet through this scission 
line. It is obtained by solving the TDGCM+GOA equation (\ref{eq:evolution0}), 
defining the current $\gras{J}(\gras{q},t)$ from the function $g(\gras{q},t)$, 
computing the instantaneous flux of that current through the scission line and, 
finally, integrating it over time; we refer to \cite{Verriere2021a} and 
references therein for further details.

\begin{figure}[!ht]
\centering
\includegraphics[width=0.9\textwidth]{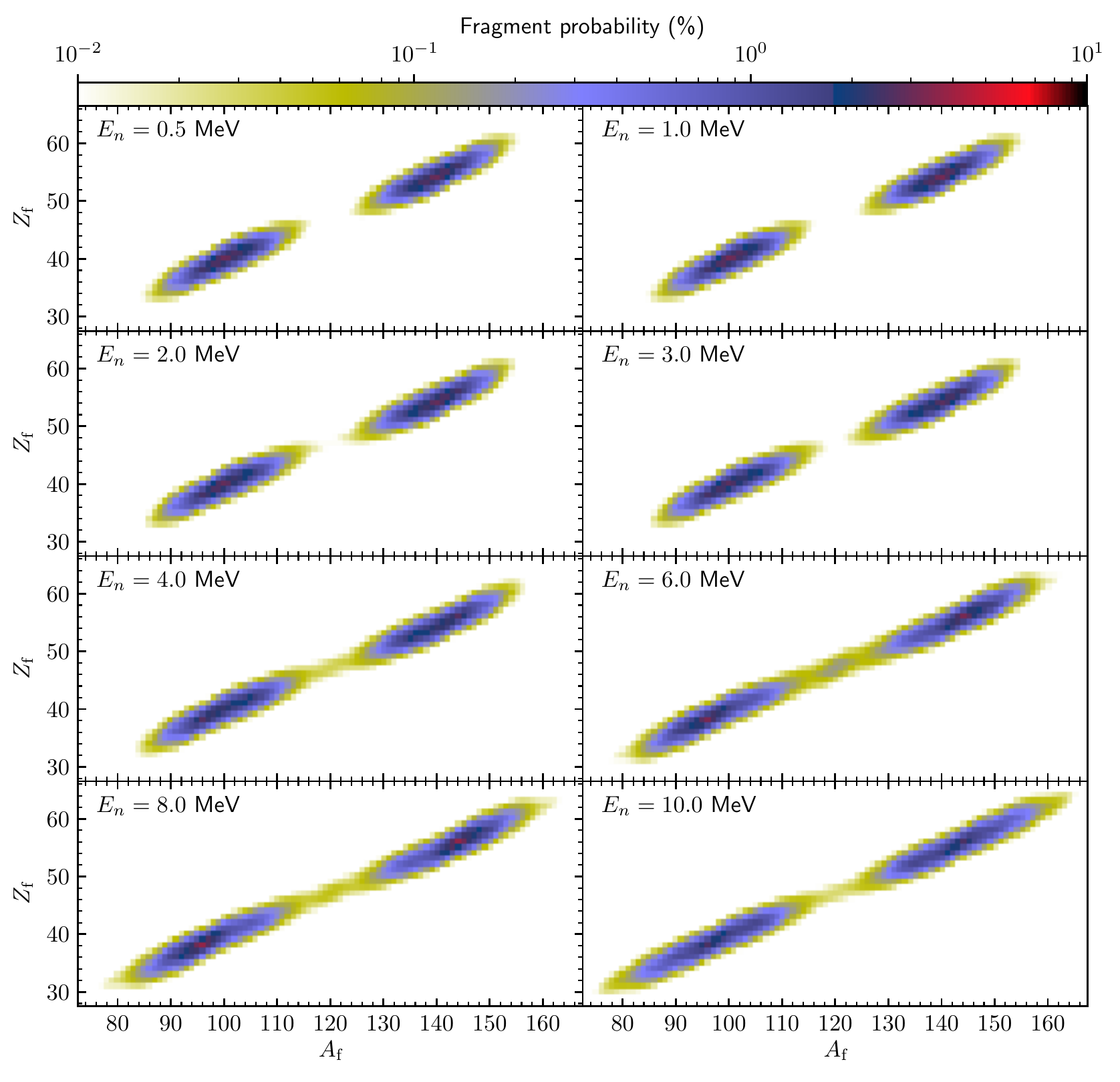}
\caption{Isotopic yields $Y(Z_f,A_f)$ for the reaction $^{239}$Pu(n,f) for 
different incoming neutron energies $E_n$. Particle number projection in both 
the fragments and the fissioning nucleus is used in each scission 
configuration. Symmetric fission becomes more probable with increasing 
excitation energy.
Figures reproduced with permission from \cite{Verriere2021a} courtesy of 
Verriere; copyright 2021 by APS.
}
\label{fig:2Dyields}
\end{figure}

In neutron-induced fission, each energy $E_n$ of the incident neutron 
corresponds to a different excitation energy of the fissioning nucleus. This 
excitation energy is used to fix the energy of the collective wave packet, 
given as 
\begin{equation}
E_{\rm coll}(t)
=
\int d\gras{q}\, g^{*}(\gras{q},t) \left[
- \frac{\hbar^2}{2} \sum_{\alpha\beta} \frac{\partial }{\partial q_{\alpha}} B_{\alpha\beta}(\gras{q}) \frac{\partial}{\partial q_{\beta}} 
+
V(\gras{q})
\right]
g(\gras{q},t) .
\end{equation}
In the TDGM+GOA, the total energy is also a constant of motion. Therefore, one 
can simulate the evolution of the fission fragment distributions with the 
energy of the incident neutron simply by changing the value of the initial 
TDGCM+GOA collective energy. The first results for both $^{239}$Pu(n,f) and 
$^{235}$U(n,f) were reported in \cite{Younes2012} and nicely reproduced the 
expected trend that symmetric fissions becomes enhanced as the neutron energy 
increases. Figure (\ref{fig:2Dyields}) shows another, more precise example of 
such calculations. Here, particle number projection in the fission fragment was 
applied to obtain a more realistic estimate of the particle number content 
(the probabilities $\mathbb{P}(Z_{\rm f},N_{\rm f} | Z,N, \gras{q}(\xi))$) of 
each scission configuration. The TDGCM+GOA equation was then solved with the 
code FELIX \cite{Regnier2018} to compute the isotopic yields in the case of 
$^{239}$Pu(n,f) reaction as a function of the incident neutron energy. The 
figure clearly shows the expected trend of enhanced symmetric fission

\section{Conclusions}

While nuclear fission has been studied for over eighty years, our understanding 
of the process remains fragmentary. Phenomenological models are invaluable for 
large-scale computations and the flexibility that their many adjustable 
parameters offer. However, improving these models require deeper insights into 
the quantum many-body features of the fission process, which only microscopic 
theories are capable of delivering. Currently, the only class of microscopic 
methods that are capable of computing actual fission observables are based on 
the general framework of the nuclear energy density functional theory. 
Essential tools include: the Hartree-Fock-Bogoliubov theory, which provides sets 
of symmetry-breaking many-body states used to generate potential energy 
surfaces and extract quantities such as fission barriers, fission paths and 
tunnelling probabilities; collective models based on the adiabatic 
approximation to large-amplitude collective motion or the generator coordinate 
method, which encode collective correlations that are especially important to 
extract the collective inertia tensor (used in the calculation of tunnelling 
probabilities) as well as to probe the entire phase space of fission; 
time-dependent density functional theory, which simulates directly fission 
events and can give rather good estimates of the initial conditions of the 
fission fragments upon their formation. 

In the short to medium term, the recent work on projection techniques as a 
tool for estimating quantities such as the number of particles or spin 
distributions of the fragments will certainly be generalized and combined with 
various time-dependent descriptions to provide complete characterization of 
fission fragments before their deexcitation. Steady increase in computational 
capabilities will allow removing several of the approximations commonly used 
when implementing EDF methods, be they the small number of degrees of freedom, 
the small size of model space, etc. Methods borrowed from the areas of machine 
learning will help quantify theoretical uncertainties in predictions of fission 
observables and generate energy functionals with better calibrated deformation 
properties. Akin to a sophisticated Lego game, several of the theoretical tools 
briefly reviewed in this chapter could, potentially, be combined with one 
another, e.g., TDGCM and TDDFT. Many of these possibilities have been discussed 
in a recent white paper by the nuclear theory community \cite{Bender2020}.

Taking the longer view, the playground of microscopic methods has so far been 
largely confined to the deformation phase of the fission process, that is, the 
evolution of the nucleus to the point of scission. The problem of fission cross 
sections, which is related to the determination of the entrance channel and the 
competition between fission and other decay modes, is still being addressed 
with semi-phenomenological methods. Switching to a microscopic approach would 
involve merging theories of nuclear structure and nuclear reactions.



\section*{Acknowledgments}

This work was supported in part by the NUCLEI SciDAC-4 collaboration 
DE-SC001822 and was performed under the auspices of the U.S.\ Department of 
Energy by Lawrence Livermore National Laboratory under Contract 
DE-AC52-07NA27344. Computing support came from the Lawrence Livermore National 
Laboratory (LLNL) Institutional Computing Grand Challenge program.

\bibliographystyle{apsrev}
\bibliography{zotero_output,books}

\end{document}